\documentclass[twocolumn]{aastex63}
\usepackage{lineno}
\usepackage{multirow}

\received{April 18, 2022}
\revised{May 27, 2022}
\accepted{\today}
\usepackage{placeins}
\FloatBarrier
\submitjournal{AJ}
\usepackage{float}
\usepackage[ansinew]{inputenc}
\usepackage[american]{babel}
\usepackage{etex}
\usepackage{stackengine}
\usepackage{csquotes}
\usepackage{ifthen}
\usepackage{url}
\usepackage{euscript}
\usepackage{multirow}
\usepackage{nameref}
\usepackage{xcolor}
\usepackage{varioref}
\usepackage{hyperref}
\usepackage{cleveref}

\shorttitle{Galactic chemical evolution of exoplanet host stars: Are high-mass planetary systems young?}

\shortauthors{C. Swastik et al.}

\graphicspath{{./}{figures/}}

\usepackage{float}
\usepackage{natbib}
\begin{document}

\title{Galactic chemical evolution of exoplanet hosting stars: Are high-mass planetary systems young?
}

\correspondingauthor{C. Swastik}
\email{swastik.chowbay@iiap.res.in}

\author[0000-0003-1371-8890]{C. Swastik}
\affiliation{Indian Institute of Astrophysics, Koramangala 2nd Block, Bangalore 560034, India}
\affiliation{Pondicherry University, R.V. Nagar, Kalapet, 605014, Puducherry, India}

\author[0000-0003-0799-969X]{Ravinder K. Banyal}
\affiliation{Indian Institute of Astrophysics, Koramangala 2nd Block, Bangalore 560034, India}

\author[0000-0002-0554-1151]{Mayank Narang}
\affiliation{Department of Astronomy and Astrophysics, Tata Institute of Fundamental Research
Homi Bhabha Road, Colaba, Mumbai 400005, India}

\author[0000-0002-3530-304X]{P. Manoj}
\affiliation{Department of Astronomy and Astrophysics, Tata Institute of Fundamental Research
Homi Bhabha Road, Colaba, Mumbai 400005, India}

\author[0000-0003-0891-8994]{T. Sivarani}
\affiliation{Indian Institute of Astrophysics, Koramangala 2nd Block, Bangalore 560034, India}

\author[0000-0003-0003-4561]{S. P. Rajaguru}
\affiliation{Indian Institute of Astrophysics, Koramangala 2nd Block, Bangalore 560034, India}

\author{Athira Unni}
\affiliation{Indian Institute of Astrophysics, Koramangala 2nd Block, Bangalore 560034, India}
\affiliation{Pondicherry University, R.V. Nagar, Kalapet, 605014, Puducherry, India}

\author{Bihan Banerjee}
\affiliation{Department of Astronomy and Astrophysics, Tata Institute of Fundamental Research
Homi Bhabha Road, Colaba, Mumbai 400005, India}

\begin{abstract}
The imprints of stellar nucleosynthesis and chemical evolution of the galaxy can be seen in different stellar populations, with older generation stars showing higher $\alpha$-element abundances while the later generations becoming enriched with iron-peak elements. The evolutionary connections and chemical characteristics of circumstellar disks, stars, and their planetary companions can be inferred by studying the interdependence of planetary and host star properties. Numerous studies in the past have confirmed that high-mass giant planets are commonly found around metal-rich stars, while the stellar hosts of low-mass planets have a wide range of metallicity. In this work, we analyzed the detailed chemical abundances for a sample of $>900$ exoplanet hosting stars drawn from different radial velocity and transit surveys. We correlate the stellar abundance trends for $\alpha$ and iron-peak elements with the planets' mass. We find the \textit{planet mass-abundance} correlation to be primarily negative for $\alpha$-elements and marginally positive or zero for the iron-peak elements, indicating that stars hosting giant planets are relatively younger. This is further validated by the age of the host stars obtained from isochrone fitting. The later enrichment of protoplanetary material with iron and iron-peak elements is also consistent with the formation of the giant planets via the core accretion process. A higher metal fraction in the protoplanetary disk is conducive to rapid core growth, thus providing a plausible route for the formation of giant planets. This study, therefore, indicates the observed trends in stellar abundances and planet mass are most likely a natural consequence of Galactic chemical evolution. 
\end{abstract}.


\section{Introduction}
When, where and how planets are formed is an actively pursued area in exoplanet science. Ever since the discovery of the first exoplanet 51 Peg by Mayor and Queloz \citep{1995Natur.378..355M} the field of exoplanets have been rapidly growing. 
With planetary census already reaching 5000 mark, it becomes statistically feasible to study the properties of the planets and their host stars to address various scientific goals. At broader level, one such goal is to understand how fundamental properties (e.g. age, mass, chemical composition, $T_{eff}$, $\log g$ etc.)  of stars hosting planets differ from stars without planets (SWP). Further insights can be gained by correlating various astrophysical properties of stellar hosts with the orbital and physical properties of the exoplanets occupying a wide parameter space.   

\begin{figure}[!t]
\centering
\includegraphics[width=1.\columnwidth]{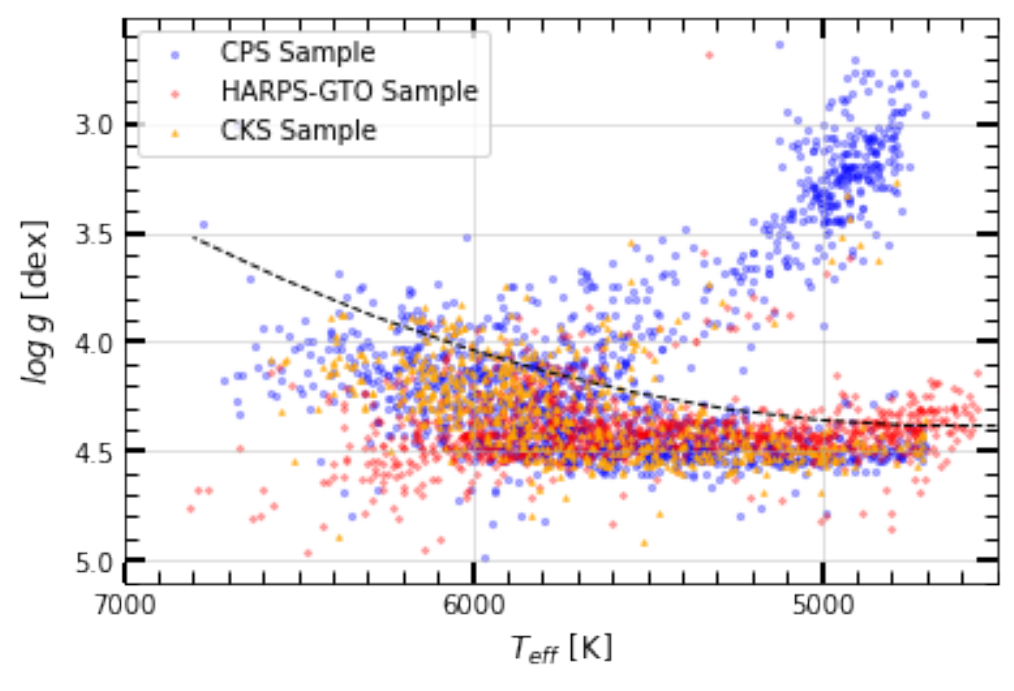}
\caption{Exoplanet host stars from the HARPS-GTO (red), CPS (blue) and CKS (yellow) surveys. The dashed black line separates the main-sequence stars lying below the line from the evolved stars lying above \citep{2018ApJS..237...38B}. In this paper, we study only the main sequence stars.}
\label{Fig1}
\end{figure}

The early detections using radial velocity techniques have shown that the occurrence of Jupiter-like planets is higher around metal-rich stars \citep[e.g.][]{1997MNRAS.285..403G, san01, fis05, 2007ARA&A..45..397U}. Subsequently, extensive spectroscopic survey of planet hosts  \citep[e.g.][]{2012MNRAS.423..122B, 2012Natur.486..375B,2014Natur.509..593B,2013ApJ...771..107E,2014ApJ...789L...3D,2015AJ....149..143F,2017AJ....154..108J,buc14, pet18, mul18,nar18, 2021AJ....161..114S} have shown that (a) host star metallicity ([Fe/H]) increases as a function of planet mass, peaking around $M_{P}\approx 4M_{J}$ and showing large scatter for the massive giant planets and brown dwarf hosts \citep{nar18,2021AJ....161..114S} (b) overall,  stars with planets tend to have higher metallicity than the stars without planets. Two main theories have been put forward to explain this metallicity excess \citep{2004A&A...426..619E}. The ``primordial" hypothesis suggested that the initial protoplanetary cloud was metal-rich, which resulted in such metal-rich hosts \citep{2004A&A...415.1153S,2008ASPC..384..292V,2010PASP..122..905J}. On the other hand, the ``self-enrichment" hypothesis attributed the high-metallicity content of the planet bearing stars to the accretion of a large amount of rocky and metal-rich planets \citep{1996Natur.380..606L,1997ApJ...491L..51L,1997MNRAS.285..403G,2001ApJ...555..801M,2001ApJ...556L..59P}. Regardless of the validity of one theory or another, the growing consensus is that a metal-rich environment plays a vital role in forming planetary systems. Particularly, the gas giants are believed to be formed from the core-accretion process which requires  fast build up of the planetary core up to  10-15$M_\oplus$. The core has to be formed quickly within a few Myr  before the gas in the disk dissipates. The metal rich protoplanetary material aids the formation of the cores followed by the accretion of the gas  to form the outer envelope.

\begin{figure}[!t]
\centering
\includegraphics[width=1\columnwidth]{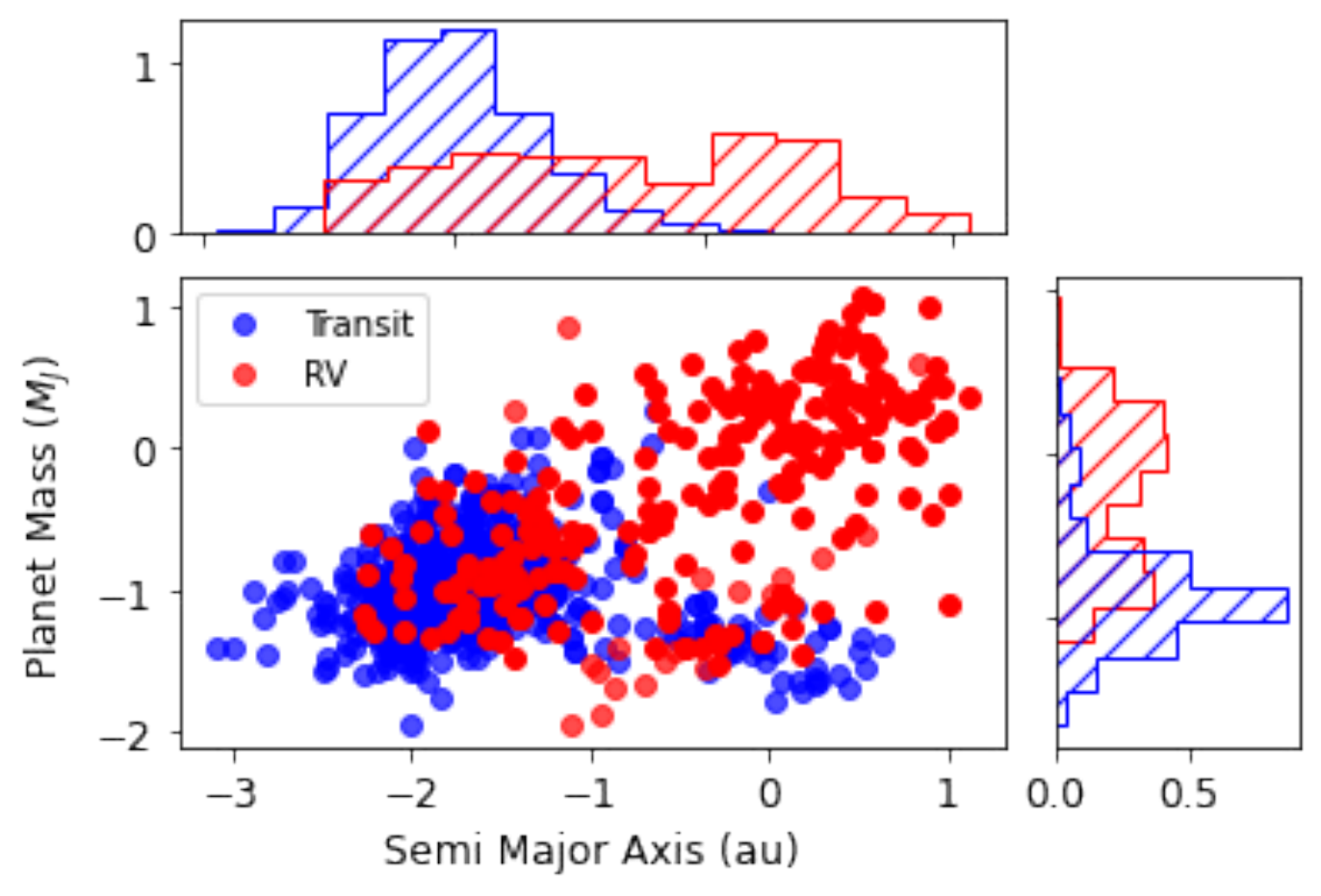}
\caption{Distribution of planet mass and semi-major axis (on log scale) for the RV and transit planets used in this study. Corresponding histogram are also shown at the top and right corner of the figure.}

\label{rvtrans}
\end{figure}

\begin{figure*}[t]
\centering
\includegraphics[width=2\columnwidth]{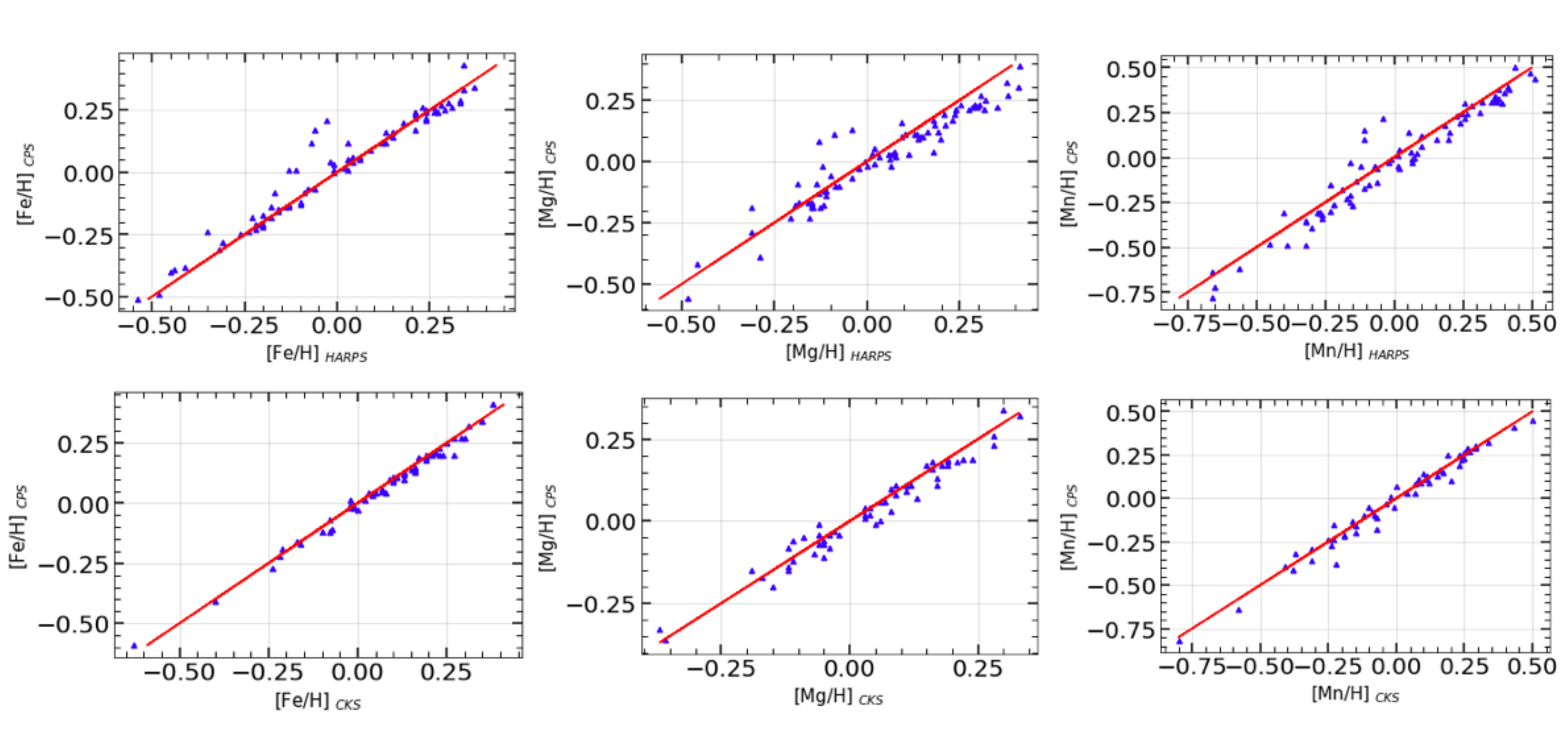}
\caption{\textbf{Top row :} Comparison of elemental abundances of Fe, Mg and Mn for stars that are common between the CPS and the HARPS-GTO sample. \textbf{Bottom row :} Comparison of Fe, Mg and Mn abundance for stars that are common between the CPS and the CKS sample. The solid red curve represents x = y line.}
\label{compare}
\end{figure*}

In the context of planet formation, most of the aforementioned studies have mainly focused on iron abundances ([Fe/H]). It is also because estimating the abundances of all the elements for a given star is not always straightforward. Although, iron is not the most abundant metal in the Universe, the optical spectra for the solar-type stars contain many prominent iron lines, making the  abundance determination easier \citep{2014A&A...569A.111B,2019Geosc...9..105A}. The iron abundance is also traditionally used as a proxy for overall metallicity of the star with the assumption that the composition of the metals changes proportionally to the iron content. However, the formation mechanism for different elements is vastly different, and their signatures do show up in the chemical composition of stars. Therefore, studying stars' detailed abundance patterns could provide further clues to the observed planet morphology and architecture.

In the past, there have been limited studies of elemental abundances, i.e., [X/Fe] for a larger sample of planet-hosting stars. For example, \citet{Brugamyer_2011,2012A&A...543A..89A,2014AJ....148...54H,2016ApJS..225...32B,2018ApJS..237...38B}, analyzed the spectra of known planetary hosts and found an overabundance of $\alpha$-elements (Mg, Si, S, Ca, Ti) for the planet-hosting stars (PHS). Similarly, \cite{2017A&A...606A..94D} and \cite{2018PASP..130i4202D} investigated the abundances of heavy elements of planet-hosting stars and found that stars with planets show an overabundance of elements such as Zn for [Fe/H] $<$ -0.1 dex. They also found most s-process elements to be under-abundant in planet-hosting stars. These studies clearly show that knowing iron content of stars alone is not sufficient, and a detailed abundance analysis is required to understand the complete picture of planet formation. The limited studies that focused on the [X/Fe] were mainly based on specific elements (such as only on $\alpha-$ or iron-peak elements). Similarly, studies such as \cite{2021arXiv211101753W} investigated the correlation between occurrence rate and chemical abundances for 10 elements for the host stars of Kepler planets. They also studied the correlation between planet radius ($R_{P}$) and abundances and detected a significant correlation between [Mn/Fe] and $R_{P}$. However these results are highly skewed towards shorter orbital period planets. A recent investigation by \cite{2022arXiv220210102T} which focused on 25 RV detected PHS and found that main-sequence giant planet-hosting stars are metal-rich compared to the low-mass planet hosting. They also found that PHS are systematically higher in $\alpha$-content than the non-hosting counterparts at the lower metallicity regime ([Fe/H]$\leq$-0.2). These studies provide a scientific  motivation for us  to investigate how the planet mass $M_{P}$ varies as a function of the abundances of different classes of elements and for a diverse sample of PHS detected both by transit and RV. Studying the [X/Fe] pattern with $M_{P}$ can also give clues about the preferred formation route for planets belonging to different mass ranges. 

\begin{deluxetable*}{ccc|ccc}[t]
\centering
\tablecaption{Table listing the samples from different surveys used in this paper.} \label{taba}
\tablewidth{0pt}
\centering
\tablehead{
& & & & \colhead{Sample after curation}\\
\colhead{Sample} &\colhead{Instrument} &\colhead{Total stars observed} & \colhead{Small planets} & \colhead{Giant planets} &\colhead{Super-jupiters}
}
 \startdata
HARPS-GTO & ESO/HARPS & 1111 & 119 & 81 & 17 \\
CKS & Keck/HIRES & 1127 & 934 & 65 & 9 \\
CPS & Keck/HIRES & 1615 & 215 & 117 & 29\\
\enddata
\tablecomments{We didn't consider the planets hosted by sub-giant stars in our sample. The final curated sample consists of only main sequence stars. }
\end{deluxetable*}

In the context of the standard galactic chemical evolution (GCE), core-collapse supernovae, mostly the Type II (SNe II), enriched the early universe with $\alpha$-elements, which also occurred on a faster time scale than Type Ia supernovae (SNe Ia) \citep{mat89,ali01, mat09, 2020ApJ...900..179K}. According to the classical view, SNe II occurs when a massive star collapses (8$M_{\odot}<$ $M_{\star}$) rapidly after the completion of its stellar burning process, which ends in an explosion. On the other hand, the most accepted model of SNe Ia involves a binary system in which at least one of the stars is a white-dwarf. The white dwarf accretes mass from its binary companion and reaches the critical mass (also known as Chandrasekhar limit), which results in thermal runaway, followed by an explosion. The SNe II produce a large amount of $\alpha$-elements and fewer iron-peak elements. The SNe Ia, on the other hand, is the major producer of iron-peak elements \citep{1993A&A...275..101E,2020A&A...634A.136C,2020ApJ...900..179K}. As a consequence of staggered progression, iron-peak elements enriched the interstellar medium (ISM) at much later stage compared to the $\alpha$-elements. Therefore, at population level, the $\alpha$ to iron-peak ratio, [$\alpha$/Fe],  in stars is a good proxy for age to probe the history of galactic chemical evolution \citep{2013A&A...560A.109H,2020A&A...634A.136C,2019A&A...624A..78D,2020ApJ...900..179K}. 

In this paper, we study the elemental abundances of a large sample of over 900 planet hosting stars with a goal to examine the role of GCE in the context of exoplanetary systems. We infer that the majority of the high-mass planetary systems ($M_{P}>0.3M_{J}$) are likely formed at later stages of the GCE, mainly after SNe Ia have sufficiently enriched the Galactic ISM with iron-peak elements. Our premise is based on the fact  that production of most elements is dictated by GCE and a heavy elements driven core-accretion mechanism is a favoured pathway for the formation of giant planets. Further motivation to this work has come from a recent study of stellar kinematic of planet hosting stars by Narang et al 2022 (under rev, AJ), suggesting that the host stars of Jupiter-type planets have a smaller velocity dispersion, which is attributed to their relatively young age. 

For this work, we use the spectroscopic abundances of planet hosting stars obtained from  three previous studies, namely, HARPS-GTO \citep{2003Msngr.114...20M,2010A&A...512A..48L,2011A&A...526A.112S}, California Kepler survey (CKS) \citep{2018ApJS..237...38B} and California planet survey (CPS) \citep{2016ApJS..225...32B}. We measured the correlation between [X/Fe] and planet mass to statistically examine if stars hosting giant planets are younger than the small planet hosts. We interpret our results in terms of GCE and mainly focus on the $\alpha$ and iron-peak elements since their formation timeline is evidently different. In our findings,  $\alpha$-elements and Eu show a strong negative correlation with planet mass, but not so significant correlation was found for the iron-peak and s-process elements. 

The rest of the paper is organized as follows. In Section~\ref{s2}, we describe our sample. In Section~\ref{xfe}, we discuss the various [X/Fe] trends as a function of planet mass. Further in Section~\ref{s4}, we compare the trends obtained in Section~\ref{xfe} and interpret our results. Finally, we give our summary and conclusions in Section~\ref{s5}.

\begin{figure*}[!t]
\centering
\includegraphics[width=2\columnwidth]{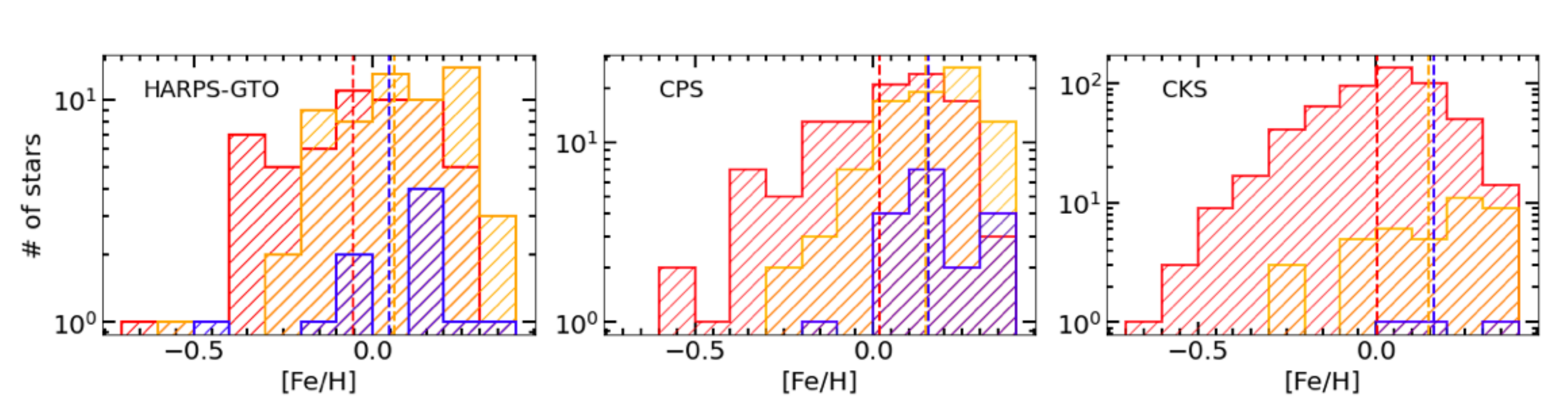}
\caption{Metallicity distribution for the HARPS-GTO, CKS and CPS sample. The colors red, yellow and blue represent small planets, giant planets and super-jupiters, respectively. The vertical lines represents the mean of the distribution.}
\label{metdist}
\end{figure*}

\section{Sample Preparation} 
\label{s2}
To study the elemental abundances of the $\alpha$, iron-peak and other elements (mainly the s-process and the r-process elements) of the exoplanet host stars, we used the data-set from three different surveys, namely, HARPS-GTO \citep{2003Msngr.114...20M,2010A&A...512A..48L,2011A&A...526A.112S}, California Planet survey (CPS) \citep{2016ApJS..225...32B}, and California Kepler survey (CKS) \citep{2018ApJS..237...38B}. The plot between T$_{eff}$ and $\log  g$ for the stars combined from the above three samples is shown in Figure~\ref{Fig1}. In this work, we analyze the main sequence stars which lie below the dashed black line shown in Figure~\ref{Fig1}. We divided the planet masses taken from NASA's exoplanet archive into three mass bins, namely, small planets (SP) [$M_{P} \leq 0.3\:M_{J}$], giant planets (GP) [$0.3\:M_{J}< M_{P} \leq 4\:M_{J}$] and super-jupiters (SJ) [$4 \:M_{J}< M_{P} \leq 13\:M_{J}$].  In this section, we briefly describe the original samples and how it was curated to obtain the final sample for our analysis. \\

\subsection{HARPS-GTO sample}
The HARPS-GTO stars used in this study come from three HARPS subsamples: \cite{2003Msngr.114...20M,2010A&A...512A..48L,2011A&A...526A.112S}.  The sample consists of 1111 F, G and K main-sequence stars  \citep{2012A&A...545A..32A} observed with HARPS, a high-resolution spectrograph (R$\sim$ 115000) at the La Silla observatory (ESO, Chile).  The HARPS-GTO sample has 163 stars with at least one companion\footnote{Data from https://exoplanetarchive.ipac.caltech.edu/} and 948 stars without any companion. The stars were extensively studied and their chemical abundances are published in a series of papers \citep{2011arXiv1109.2497M,2011A&A...535L..11A,2012A&A...545A..32A,2017A&A...599A..96S,2015A&A...576A..89B,2016A&A...591A..69S,2018PASP..130i4202D,2019A&A...624A..78D,2017A&A...606A..94D,2021arXiv210904844D,2020A&A...634A.136C}. The technique employed to obtain the elemental abundances is  mostly the equivalent width method. Initial study for the HARPS sample was done for elements with A $<$ 29 by \cite{2012A&A...545A..32A} which focused mainly on chemical separation of thin and thick disk stars. The study also showed an over-abundance of all the elements ([X/H]) for the giant planet hosts. However, no trends for [X/Fe] with planet mass were studied. We took the elemental abundance of eight elements (Mg, Si, Ca, Ti, Cr, Ni, Co, Mn) from \cite{2012A&A...545A..32A} for our analysis.

For the neutron-capture elements, a separate study was conducted for the HARPS-GTO sample by \cite{2017A&A...606A..94D} and \cite{2018PASP..130i4202D}. For the s-process such as Ba, Sr, Ce, and Zr, it is found that planet-hosting stars are under-abundant compared to stars without planetary companion. These results are significant as they throw light on how s-process elemental abundances plays a role in distinguishing stars with and without planets. However,  the stellar abundance as a function of planet mass were not studied in detail in these papers. Thus, we took the elemental abundances ([X/Fe]) for two iron-peak elements (Cu, Zn) and seven neutron capture elements (Sr, Y, Zr, Ba, Ce, Nd and Eu) from \cite{2017A&A...606A..94D} and combined it with the eight elements from \cite{2012A&A...545A..32A} to study the trends of $\alpha$, iron-peak, s and r-process elements as a function of planet mass.

\subsection{California planet survey (CPS)}
\label{cps}
The abundance of stars in CPS is taken from \cite{2016ApJS..225...32B}. The sample consists of 1615 F, G, K and M stars which were observed using HIRES spectrograph (R$\sim$ 70000) on the KECK I telescope as a part of radial velocity planet search program \citep{2010PASP..122..149J,2010ApJ...721.1467H,2011ApJ...730...93W,2016ApJS..225...32B}. These stars were observed in the red configuration of HIRES without iodine cells in the beam path. We used the abundances of nine elements (Mg, Si, Ca, Ti, Cr, Mn, Fe, Ni, and Y) for our analysis from the CPS sample, which was obtained using the synthetic spectral fitting \citep{2016ApJS..225...32B} (SME; \cite{2017A&A...597A..16P}). To extract the planet-hosting stars from the sample, we cross-matched the CPS catalogue with NASA exoplanet archive \citep{2013PASP..125..989A,https://doi.org/10.26133/nea12} with a search radius of 3$\arcsec$ (see \cite{2020AJ....159..194V} for details)\footnote{We used a larger search radius initially but found that search radius of 3$\arcsec$ was sufficient in this case to extract all the planet-hosting stars. We also double-checked them with other parameters such as Simbard name to verify if they are truly planet-hosting stars.} and found that there are 227 stars hosting 361 planets. 

\subsection{California Kepler Survey (CKS)}
\label{cks}
The CKS sample used in this study comes from \cite{2018ApJS..237...38B}. It consists of 1127 stars which are Kepler objects of interest (KOI). The CKS sample primarily consists of KOIs with the magnitude in the Kepler band $K_{P}\leqslant14.2$ \citep{2011ApJ...736...19B,2017AJ....154..108J,2017AJ....154..107P}. The CKS KOIs used in this study were observed using the same instrumental configuration as that of the CPS host stars described in Section~\ref{cps}. For our analysis, we took the elemental abundances for nine elements (Mg, Si, Ca, Ti, Cr, Mn, Fe, Ni, and Y) for the KOIs from \cite{2018ApJS..237...38B}, which used synthetic spectral fitting similar to the CPS host stars. We crossmatched the CKS data used in this study with NASA exoplanet archive \citep{2013PASP..125..989A,https://doi.org/10.26133/nea12} with a search radius of 3$\arcsec$ (same as done for CPS) and found a total of 600 stars hosting at least one planet. The remainder of the sample consists of planetary candidates, false positives and stars without planets (see the \href{https://exoplanetarchive.ipac.caltech.edu/cgi-bin/TblView/nph-tblView?app=ExoTbls&config=fpwg}{kepler false-positive table} for details). For our analysis, we have only considered the main sequence stars from the CKS sample, which hosts confirmed planets. 

\startlongtable
\begin{deluxetable*} {lllllllllllll} 
\centering
\tablecaption{Key parameters of exoplanet host stars  used in this study.} \label{tabb}
\tablewidth{0pt}
\centering
\tablehead{\colhead{Star ID} &\colhead{Planet name} &\colhead{RA} & \colhead{DEC} & \colhead{Survey} &\colhead{$M_{P}(M_{J})$}&\colhead{{[}Fe/H{]}} &\colhead{{[}Mg/Fe{]} }&\colhead{{[}Si/Fe{]}} &\colhead{{[}Ca/Fe{]} }}
 \startdata
HD 100777 & HD 100777 b & 173.9646761 & -4.7556922 & HARPS-GTO & 1.03 & 0.25 & 0.04 & 0.05 & -0.035  \\
HD 10180 & HD 10180 c & 24.4732364 & -60.5115264 & HARPS-GTO & 0.04122 & 0.08 & 0.04 & 0.02 & 0.012  \\
HD 10180 & HD 10180 d & 24.4732364 & -60.5115264 & HARPS-GTO & 0.03697 & 0.08 & 0.04 & 0.02 & 0.012  \\
... & & & & & & & & &
\enddata
\tablecomments{The entire table is available in machine readable format. For brevity, first 3 rows and 10 columns are only shown here.}
\end{deluxetable*}

\subsection{Planet Mass}
The data for the planet mass was mostly obtained from the NASA exoplanet archive \citep{2013PASP..125..989A,https://doi.org/10.26133/nea12}. For 24 planets, the masses were taken from the exoplanet.eu catalogue for which the mass was not available in the NASA exoplanet archive. For the transiting planets in CKS sample, the mass is derived from the \textit{mass-radius} relation given by \cite{che17}. For planets detected by radial velocity (RV) in the HARPS-GTO and CPS, the minimum mass ($M_{P}$) was used. For the giant planets in the CKS sample for which the RV follow-up observations were done, the actual mass derived from the RV analysis was used. The distribution of planet mass and orbital distance for our sample is shown in Figure~\ref{rvtrans}.

\begin{figure*}[!t]
\includegraphics[width=2\columnwidth]{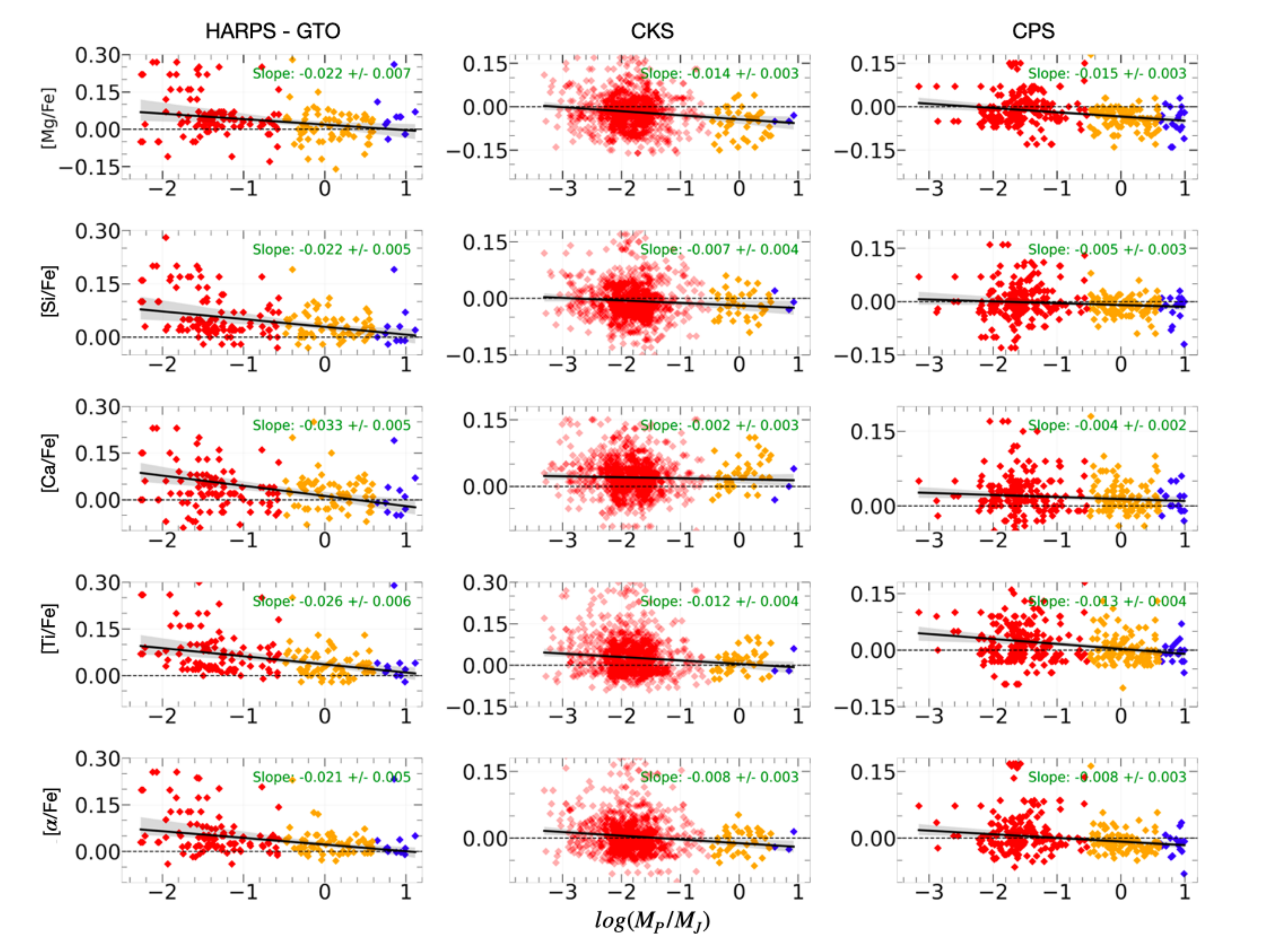}
\caption{Observed trends for $\alpha$-element abundances of host stars and planet mass for the HARPS-GTO, CKS and CPS sample. The colors red, yellow and blue represent small planets, giant planets and super-jupiters, respectively. The black line shows the Huber regression fit and the grey shaded region represents the 95 percentile confidence interval. Slope value for the best fit line is shown in each panel.  The last row is the arithmetic mean, of the $\alpha$-element abundance from the above four rows. }
\label{al}
\end{figure*}

\subsection{Abundance Comparison }
The elemental abundances derived by different techniques suffer from systematic biases \citep{bla19}. The abundances for the host stars of CPS and CKS are derived by synthetic spectral fitting using spectroscopy made easy (SME; \cite{2017A&A...597A..16P}). On the other hand, the abundances of the HARPS-GTO sample were primarily determined using the equivalent width method using MOOG \citep{1973ApJ...184..839S}. We wanted to compare if the elemental abundances obtained by the two different groups have any significant offset or scatter amongst them. We found 79 stars common between HARPS-GTO and CPS samples while 56 stars common between CPS and CKS samples. Figure~\ref{compare} shows the abundance comparison between the HARPS-GTO vs CPS and CKS vs CPS sample for three different elements;  Fe, which is used as a proxy for overall metallicity; Mg, which is an $\alpha$-peak element and Mn, which is a iron-peak element. Most stars in CPS and CKS samples were not only observed using the same telescope and instrumental setup, but also analyzed using the same techniques and thus show a less spread about x = y line in the Figure~\ref{compare}.  Barring few outliers, our analysis indicates that overall abundances determined in three samples using different methods are largely consistent. Since we didn't find any significant offset or scatter between our samples,  no correction was made for further study.

\subsection{Final sample}
The distribution of $T_{eff}$ and $\log g$ for the original HARPS-GTO, CPS and CKS samples is shown in Figure~\ref{Fig1}. The original sample include many evolved stars, mostly sub-giants. In this study, we have restricted our analysis to the main sequence stars, since, for the sub-giants, it is difficult to account for NLTE and evolutionary effects which can cause  mixing in the photospheric abundances. Following the procedure of \cite{2018ApJS..237...38B}, we selected stars below the black dashed line (see Figure~\ref{Fig1}), which represents the main sequence stars. Thus, our final data consists of 217 planets hosted by 141 stars from the HARPS-GTO sample, 600 stars hosting 1008 planets from the CKS sample, and 227 stars hosting 361 planets in the CPS sample. A detailed description of our final sample is given in the Table~\ref{taba}.

\begin{figure*}[!t]
\includegraphics[width=2\columnwidth]{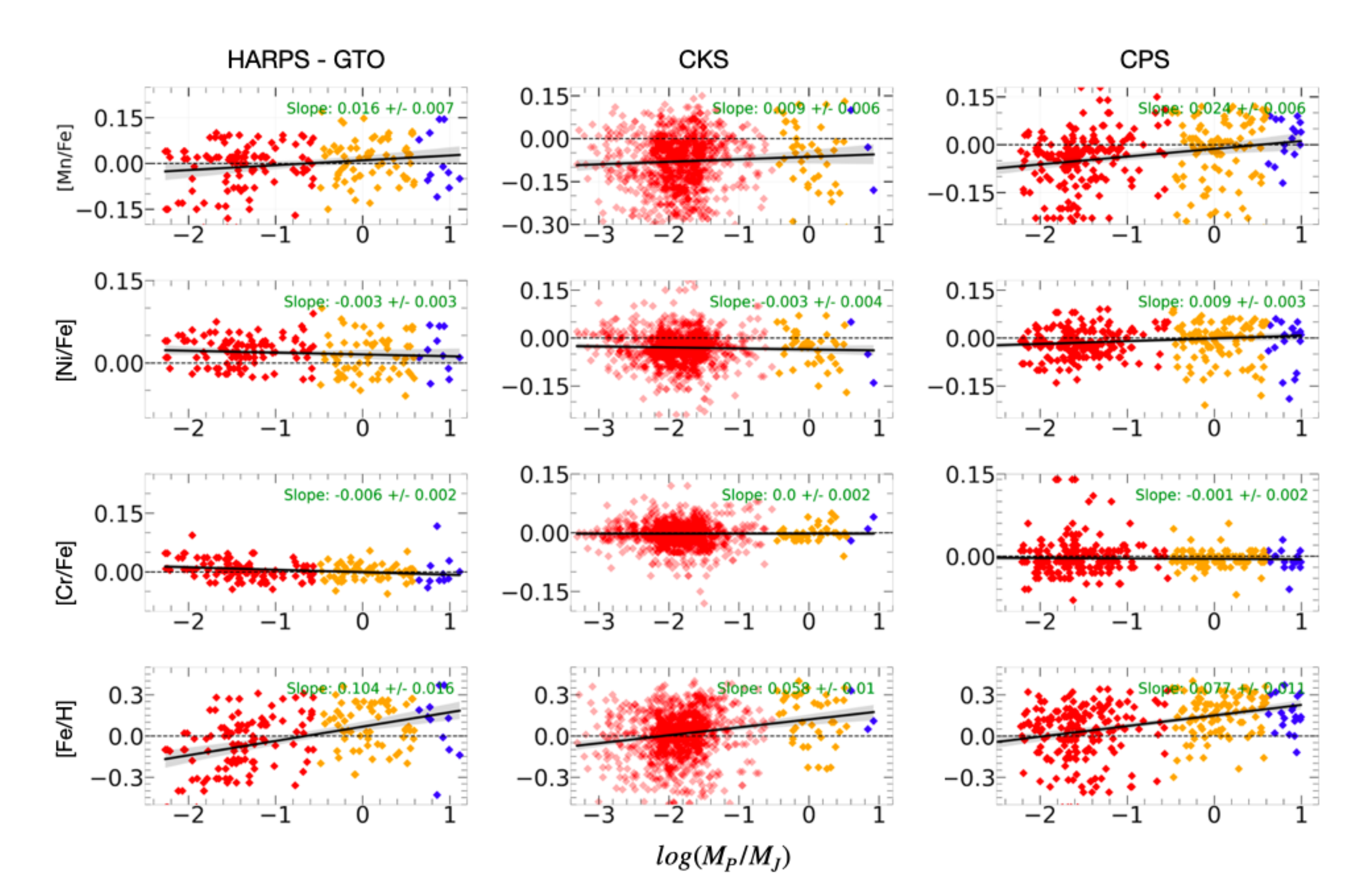}
\caption{Host star chemical abundances for iron-peak (Mn, Cr and Ni) elements and metallicity ([Fe/H]) as a function of planet mass for the  HARPS-GTO, CKS and CPS sample. Symbols and the colors are same as that of Figure~\ref{al}.}
\label{ir}
\end{figure*}

\begin{figure*}[!t]
\includegraphics[width=2\columnwidth]{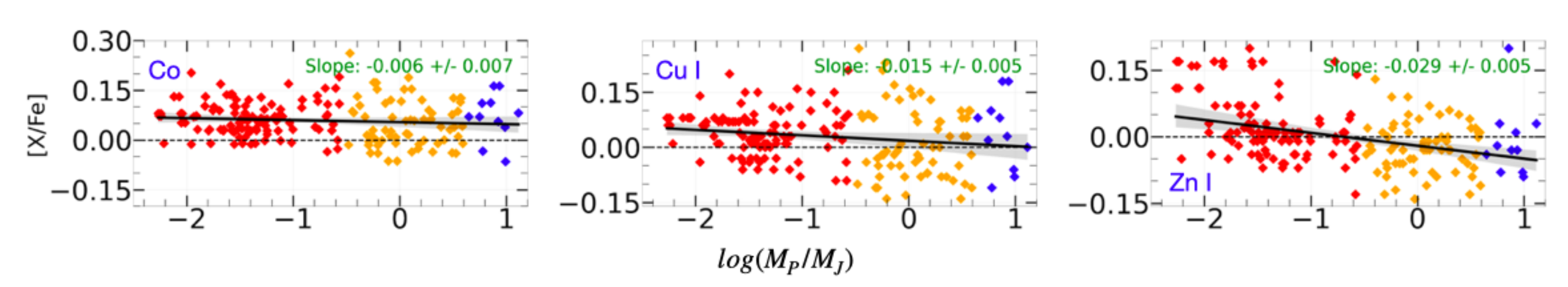}
\caption{Stellar abundances for iron-peak elements (Co, Cu and Zn) as a function of planet mass exclusively for HARPS-GTO sample. The Co, Cu and Zn abundances  were not available for CKS and CPS samples. Once again, the color scheme and the black line representation is same as that of Figure~\ref{al}.}
\label{ir1}
\end{figure*}

\section{Analysis and Results}
\label{xfe}
One of the goals of this work is to examine the correlation between the abundance of the host stars with their planet mass and how it relates to the chemical evolution of elements in the galaxy. The knowledge of different elements produced in various stages of GCE can help us understand the observed trends between [X/Fe] and planet mass. In fact, such trends would be indicative of timescale when planets of different masses were formed. We used regression analysis and Spearman's coefficients  to study the correlation between planet mass and chemical abundances [X/Fe] of the stellar hosts. In standard linear regression, the presence of outliers can significantly influence the least-squares fit which approximates the underlying trends between the parameters of interest.  We, therefore, used the Huber regression model, which is a robust approach to produce a `weighted' regression line that is less sensitive to outliers. Furthermore, to keep our regression analysis simple, we have not included stars hosting multiple planets belonging to SP, GP and/or SJ category. The list of multiplanetary systems comprises: 14 stars hosting 34 planets in HARPS-GTO; 24 stars hosting 70 planets in CKS; and 16 stars hosting 42 planets in CPS sample. The selected planet hosting stars and associated stellar and planetary parameters are given in Table~\ref{tabb}. Figure~\ref{metdist} shows the histogram of metallicity distribution of our sample. Clearly, the massive planets ($>0.3M_{J}$) are mainly hosted by metal-rich stars, while for the smaller planets ($<0.3M_{J}$) there is no specific preference of in terms of metallicity of the host stars. The regression trends for various elements are presented in the following subsections.

\begin{figure*}[thb]
\centering
\includegraphics[width=2\columnwidth]{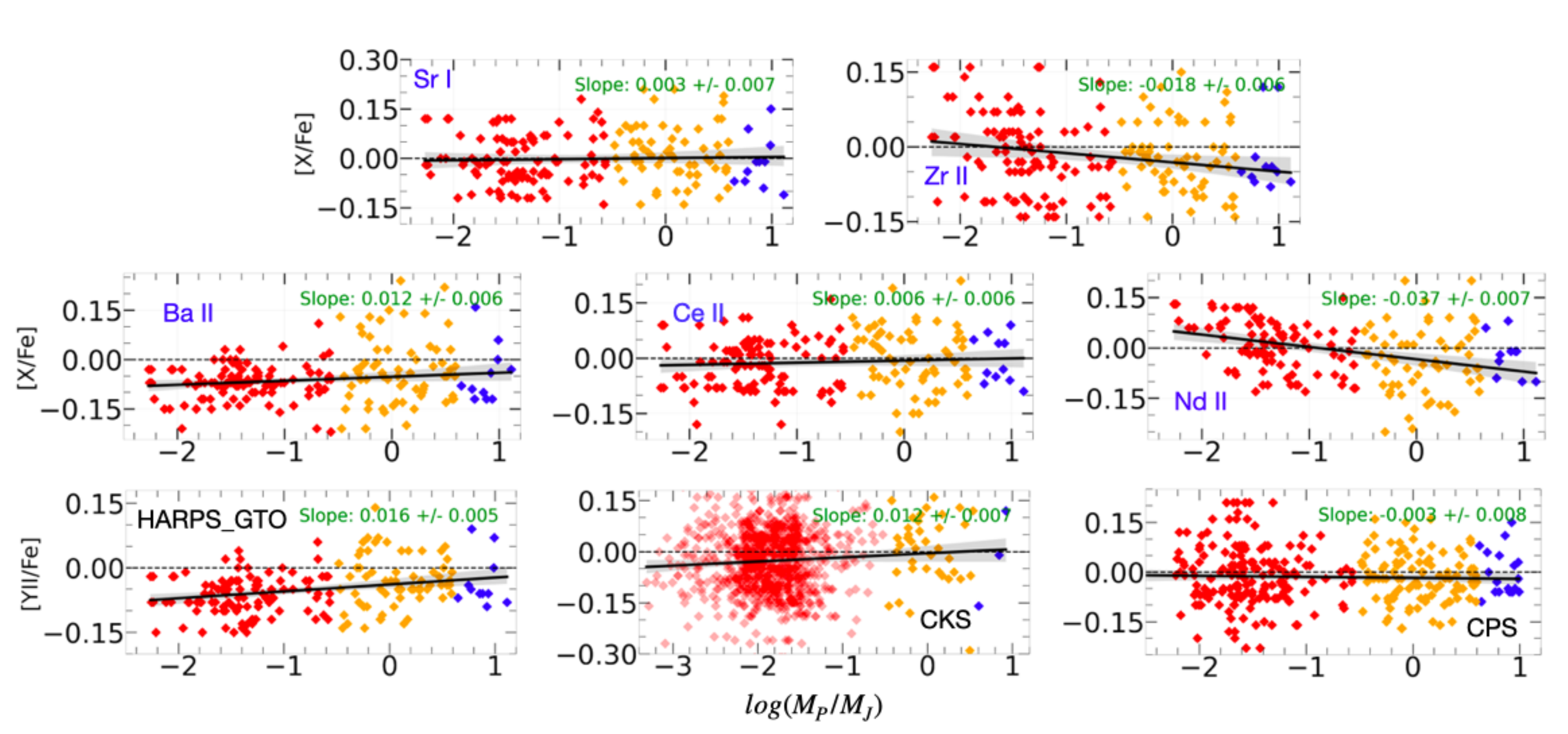}
\caption{\textbf{Top row :} Host star chemical abundances for light s-process elements as a function of planet mass for the HARPS-GTO sample. \textbf{Middle row :} Host star chemical abundances for heavy s-process elements as a function of planet mass for the HARPS-GTO sample. The color scheme is same as that of Figure~\ref{al}. \textbf{Bottom row :} Light s process element (Y) abundances as function of planet mass for HARPS-GTO, CKS and CPS sample.} 
\label{lsp}
\end{figure*}

\subsection{$\alpha$-elements}
\label{salpha}
The  $\alpha$-element abundances of the planet hosting stars can be a proxy to the age of the stars \citep{2019A&A...624A..78D}. A significant contribution of $\alpha$-elements comes from SNe II. In this paper, we examine the abundance pattern for four common $\alpha$-elements (Mg, Si, Ca and Ti) that were studied in the HARPS, CPS and CKS samples. Figure~\ref{al} shows the variation of $\alpha$-abundances of host stars with the mass of their planetary companions. The last row in the Figure~\ref{al} represents the mean abundance of all four $\alpha$-elements in each sample. The uncertainties associated with the individual abundance measurements are about 0.02-0.05 dex. In the HARPS-GTO case, we find a clear negative correlation for all the $\alpha$-element abundances with the planet mass.  In the case of the CKS sample, which is dominated by small planets, the correlation is weaker compared to the HARPS-GTO sample. The CPS sample also shows significant negative correlation for all the elements. A strong (weak)  correlation implies large (small) Spearman's rank coefficient and small (large) p-value as shown in Table~\ref{tab1}. 

Clearly, there is an overall decline of [$\alpha$/Fe] with increasing planet mass in all the three samples. This can be understood as gradual enrichment of ISM with iron produced in SN~Ia, and not necessarily the decline of [$\alpha$/H], which can also be seen in the Appendix~\ref{app}, Figure~\ref{alfe}. Interestingly, the regression analysis done separately for multiplanetary systems (Appendix~\ref{apb}, Figure~\ref{mulp}) does not show any significant correlation of [$\alpha$/Fe] with the planet mass. Additionally, the overall $\alpha$-element abundances for this sample is also found to be lower across three mass-bins. 

\subsection{Iron-peak elements}
The significant contribution of iron-peak elements comes from the Type-Ia supernovae, which occurred at the later stages of GCE compared to SNe II. Again, we analyzed the iron-peak abundance trends for three elements (Cr, Mn, Ni) common for HARPS-GTO, CKS and CPS samples and three elements (Co, Cu, Zn) exclusively from HARPS-GTO samples. Figure~\ref{ir} (except the last row) and \ref {ir1} shows the iron-peak abundances trends as a function of planet mass. We also find a positive correlation for the Mn abundance with planet mass for all three samples. On the other hand, the abundance of Zn shows behaviour similar to $\alpha$-peak elements (a strong decreasing trend). This is likely because Zn is also synthesised in core-collapse supernovae. Therefore, we see the same effect of GCE in Zn as we see in the $\alpha$-elements \citep{2020ApJ...900..179K}. For Co, Ni and Cu, we don't see any significant abundances trends with planet mass. In the case of Cr, we see a negative trend for the HARPS-GTO sample, but we don't see any trend for CKS and CPS samples. The last row of Figure~\ref{ir} shows  the increasing trend of stellar metalicity ([Fe/H]) with planet mass, which is a well established result reported in many similar studies \citep{fis05,nar18}.

Overall, the iron-peak elements don't show any significant correlation with planet mass (as listed in Table~\ref{tab1}) except for Mn and Zn. Also, the enrichment of Fe-peak elements with Fe is either increasing (for Mn, Co, Ni and Cu I) or zero (Cr), as seen in Figure~\ref{irfe} which is in sharp contrast to the trends for $\alpha$ elements.

\label{sipk}
\subsection{Heavy-elements}
Stellar fusion alone cannot produce elements heavier than iron. Most of the heavy-elements (A$>$30) are formed by neutron capture process which can be broadly classified into slow- and rapid- process. The slow-process (s-process) takes place when the density of neutrons is low ($n_{n}\sim10^{8}$ cm$^{-3}$), and the successive captures of neutrons happen at a longer time scale ($\sim10^{3}-10^{4}$ years) \citep{doi:10.1146/annurev.astro.43.072103.150600,karakas_lattanzio_2014,2018ARNPS..68..237F,2020ApJ...900..179K}. If the nuclei are unstable, then a $\beta$-decay will occur, transforming neutron to protons (thus increasing atomic number). In the case of rapid-process (r-process), the density of neutrons is higher ($n_{n}>10^{22}$ cm$^{-3}$), therefore, the time scale is much shorter ($\sim$few milliseconds to seconds) between the subsequent neutron captures compared to s-process ($\sim10^{3}-10^{4}$ years). Also since the r-process time scale is much shorter than the $\beta$-decay time scales \citep{1992A&A...258..357B,2021RvMP...93a5002C}, the r-process happens much faster. The GCE trends for the various heavy elements with Fe are shown in Figure~\ref{lspfe}. For all the heavy elements, in the region [Fe/H]$>$-0.5, we find a gradual decrease in [heavy elements/Fe] abundances with Fe-enrichment.

\begin{deluxetable*}{ccccccc}[t]
\tablecaption{Spearman's rank correlation coefficient $\rho$ values obtained between elemental abundance [X/Fe] and planet mass for HARPS-GTO, CKS and CPS sample. The values in the parenthesis represents the p-values associated with the correlation.} \label{tab1}
\tablewidth{0pt}
\centering
\tablehead{
\colhead{Category} &\colhead{Element} &\colhead{Atomic number} & \nocolhead{Name}  & \colhead{$\rho$ (p-value)} &\nocolhead{Name} 
\\
\cline{4-6}
\nocolhead{Name}  & \nocolhead{Name}  & \nocolhead{Name}  &\colhead{HARPS}
&\colhead{CKS} &\colhead{CPS}
}
 \startdata
$\alpha$-elements & Mg & 12 & -0.27 (1.93$\times$10$^{-4}$)& -0.10 (2.3$\times$10$^{-3}$)& -0.28 (5.71$\times$10$^{-7}$) \\
 & Si & 14 & -0.39 (4.02$\times$10$^{-8}$)& -0.10 (3.05$\times$10$^{-3}$) & -0.10 (4.01$\times$10$^{-2}$) \\
 & Ca & 20 & -0.42 (1.14$\times$10$^{-9}$)& -0.06 (8.02$\times$10$^{-2}$) & -0.13 (1.23$\times$10$^{-2}$)\\
 & Ti & 22 & -0.45 (1.69$\times$10$^{-10}$) & -0.13 (7.18$\times$10$^{-5}$)& -0.25 (2.84$\times$10$^{-6}$) \\
 & $\alpha$-avg & -- & -0.45 (5.92$\times$10$^{-9}$) & -0.13 (5.661$\times$10$^{-5}$)& -0.25 (6.42$\times$10$^{-6}$)\\
\hline
Iron-peak & Cr & 24 & -0.22 (2.91$\times$10$^{-4}$)& -0.01 (5.55$\times$10$^{-1}$)& 0.00 (9.72$\times$10$^{-1}$) \\
 & Mn & 25 & 0.17 (2.28$\times$10$^{-2}$)& 0.08 (3.52$\times$10$^{-2}$) & 0.26 (9.37$\times$10$^{-7}$)\\
 & Co & 27 & -0.08 (1.37$\times$10$^{-1}$)& --&--\\
 & Ni & 28 & -0.08 (2.08$\times$10$^{-1}$) & -0.01 (6.47$\times$10$^{-1}$) & 0.12 (4.44$\times$10$^{-3}$)\\
 & Cu & 29 & -0.14 (4.03$\times$10$^{-2}$) & -- & -- \\
 & Zn & 30 & -0.39 (3.21$\times$10$^{-8}$) & -- &--\\
 \hline
Light s-process  & Sr I& 38 & 0.03 (0.63) & -- & -- \\
 & Y II& 39 & 0.27 (6.99$\times$10$^{-5}$)& 0.01 (5.63$\times$10$^{-1}$)& -0.04 (3.11$\times$10$^{-1}$) \\
 & Zr II & 40 & -0.21 (3.97$\times$10$^{-3}$) & -- & --\\
 \hline
Heavy s-process  & Ba II& 56 & 0.17 (2.68$\times$10$^{-2}$)& -- & --\\
 & Ce II& 58 & 0.06 (3.16$\times$10$^{-1}$) & --&--\\
 & Nd II& 60 & -0.36 (1.09$\times$10$^{-6}$)& -- &--\\
\hline
r-process  & Eu II & 63 & -0.37 (1.07$\times$10$^{-6}$)& --& -- \\
\enddata

\end{deluxetable*}

\begin{figure}[!b]
\includegraphics[width=0.9\columnwidth]{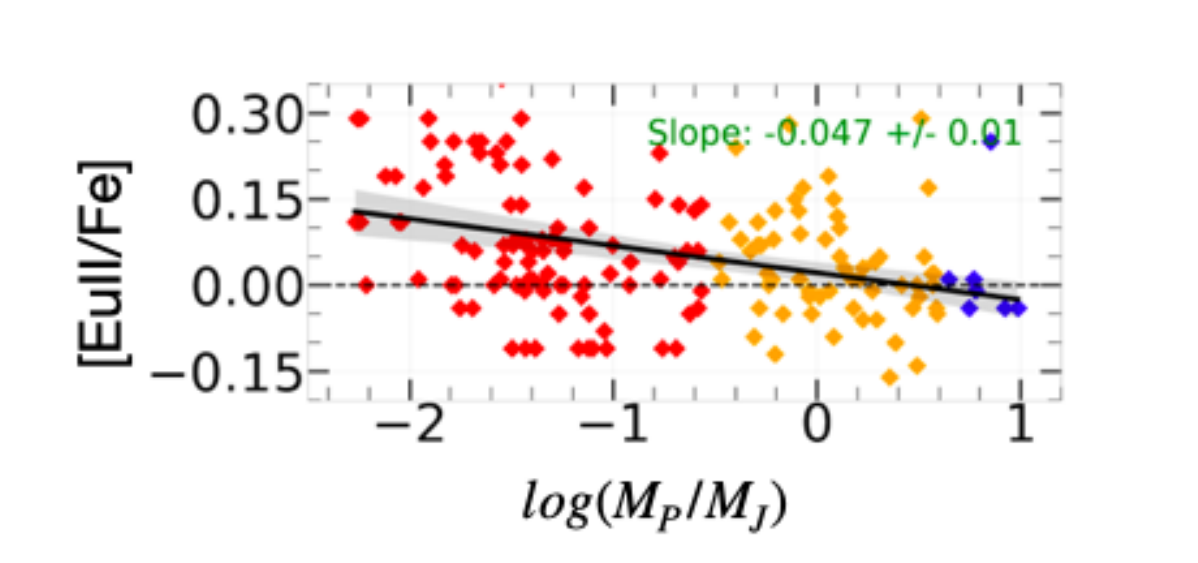}
\caption{Variation of Eu abundance (r-process element)  as a function of planet mass for the HARPS-GTO sample. The color scheme is same as that of Figure~\ref{al}.}
\label{eu}
\end{figure}

\subsubsection{Light s-process elements}
The major production site for the s-process elements is in the He intershell of the asymptotic giant branch (AGB) stars \citep{1992A&A...258..357B,2021A&A...649A..49G,2022A&A...657A..50G}. The s-process elements are further categorized based on their atomic masses. Here, we studied the abundances for three light s-process elements (Y, Sr, Zr). The top row and bottom row in the Figure~\ref{lsp} shows the light s-process abundances trends as a function of planet mass. For YII and SrI we don't see any significant correlation in our samples, whereas for ZrII, we find a negative trend with planet mass as pointed in Table~\ref{tab1}.

\subsubsection{Heavy s-process elements}
The three heavy s-process elements analyzed in this work are Ba, Ce and Nd. The middle row of the Figure~\ref{lsp} shows the heavy s-process abundances trends as a function of planet mass. We find that the correlation between BaII and CeII abundances and planet mass is weak. On the other hand, Nd shows are a strong negative trend as planet mass increases. The behaviour of Nd resembles $\alpha$-elements. 

\subsubsection{r-process elements}
Although the formation mechanism of r-process elements is a field of active research, with the recent observations of kilonova GW170817, it is possible to explain the Eu abundances solely from neutron star merger models \citep{2019MNRAS.483.4397V}. The only pure r-process element known and is studied here is Eu from the HARPS-GTO sample. Figure~\ref{eu} shows the strong negative trend of Eu with planet mass which looks similar to the $\alpha$-elements.

\section{Discussion}
\label{s4}
\subsection{$\alpha$-elements: proxy to planet mass and age}
\label{D1}

The $\alpha$-elements primarily formed by SNe II, which happened at the earlier stages of the GCE, while the iron-peak elements believed to have formed during the SNe Ia, occurring at the later stages of GCE. Relative to iron, the abundances of $\alpha$-elements and those formed mostly by SNe II in general increases with the age of the star \citep{2015A&A...579A..52N,2018ApJ...865...68B,2018A&A...619A.125A,2018MNRAS.477.2326F,2019A&A...624A..19B,2019A&A...624A..78D}. From Figure~\ref{al}, we see that [$\alpha$/Fe] and  planet mass have negative slope. In addition, the low-mass planets hosts show larger [$\alpha$/Fe] dispersion compared to the parent stars of Jupiter and super-jupiters (see Section~\ref{AGE} for further discussion). One plausible interpretation of such trends is that the low-mass rocky planets have been forming around all generation of stars (old as well as young), while the high-mass giant planets likely  formed around younger stars when ISM was sufficiently enriched with iron-peak elements. Same reasoning must apply to the multiplanetary systems hosting at least one low-mass and one high-mass planet such as Jupiter or super-jupiters. As shown in Appendix~\ref{apb}, Figure~\ref{mulp},  the slope between $\alpha$-element abundance and planet mass is nearly an order of magnitude smaller compared to the corresponding slopes in Figure~\ref{al}. This implies, multiplanetary systems accompanying at least one high-mass planet are clearly $\alpha$-deficient and therefore, younger. This wouldn't be the case for multiplanetary systems hosting only the small planets.

Since the iron-peak elements are formed at later stages of GCE, it suggests a similar formation timeline for the hosts of giant planets and possibly super-jupiters, if core accretion was dominant mechanism. Moreover, the abundance of iron-peak elements scales in same way as the abundance of iron [Fe/H].  Thus, the trends for iron-peak elements with planet mass is nearly positive or zero, as expected except Zn, which shows a strong negative trend similar to $\alpha$-elements. The anomalous behavior of Zn is also seen in several studies  \citep[e.g.][]{2004MmSAI..75..741B,2017A&A...600A..22M,2019A&A...624A..78D}. Zn is found to increase with age, as it is also synthesised in core-collapse supernovae, and thus follows the GCE trends similar to $\alpha$-elements \citep{2020ApJ...900..179K}. We also see a positive trend for Mn as planet mass increases. Mn is produced mostly in SNe Ia \citep{1997NuPhA.621..467N,2006ApJ...653.1145K} and this trend indicates that statistically massive planet hosts are Mn-rich, and the presence of Mn in the host star may be crucial in the formation of giant planets. Interestingly, the GCE effect is also  strong for [Mn/Fe] even for the field stars as evident from Figure~\ref{irfe}. However that alone cannot explain a relatively large slope of [Mn/Fe] versus planet mass among iron-peak elements seen in Figure~\ref{ir}. The yield for Ni are quite similar during SNe Ia and SNe II \citep{2013ARA&A..51..457N,2017A&A...600A..22M}, thus, we expect a flat trend with planet mass which is also seen observationally. In line with GCE, the absence of any significant trend in  iron-peak elements with planet mass independently suggests that production of most iron-peak elements co-evolved with Fe. But more importantly, the later enrichment of the ISM with Fe and iron-peak elements, as the trends indicate, could be an important ingredient for the formation of high-mass planets.

\begin{figure}[!t]
\includegraphics[width=0.9\columnwidth]{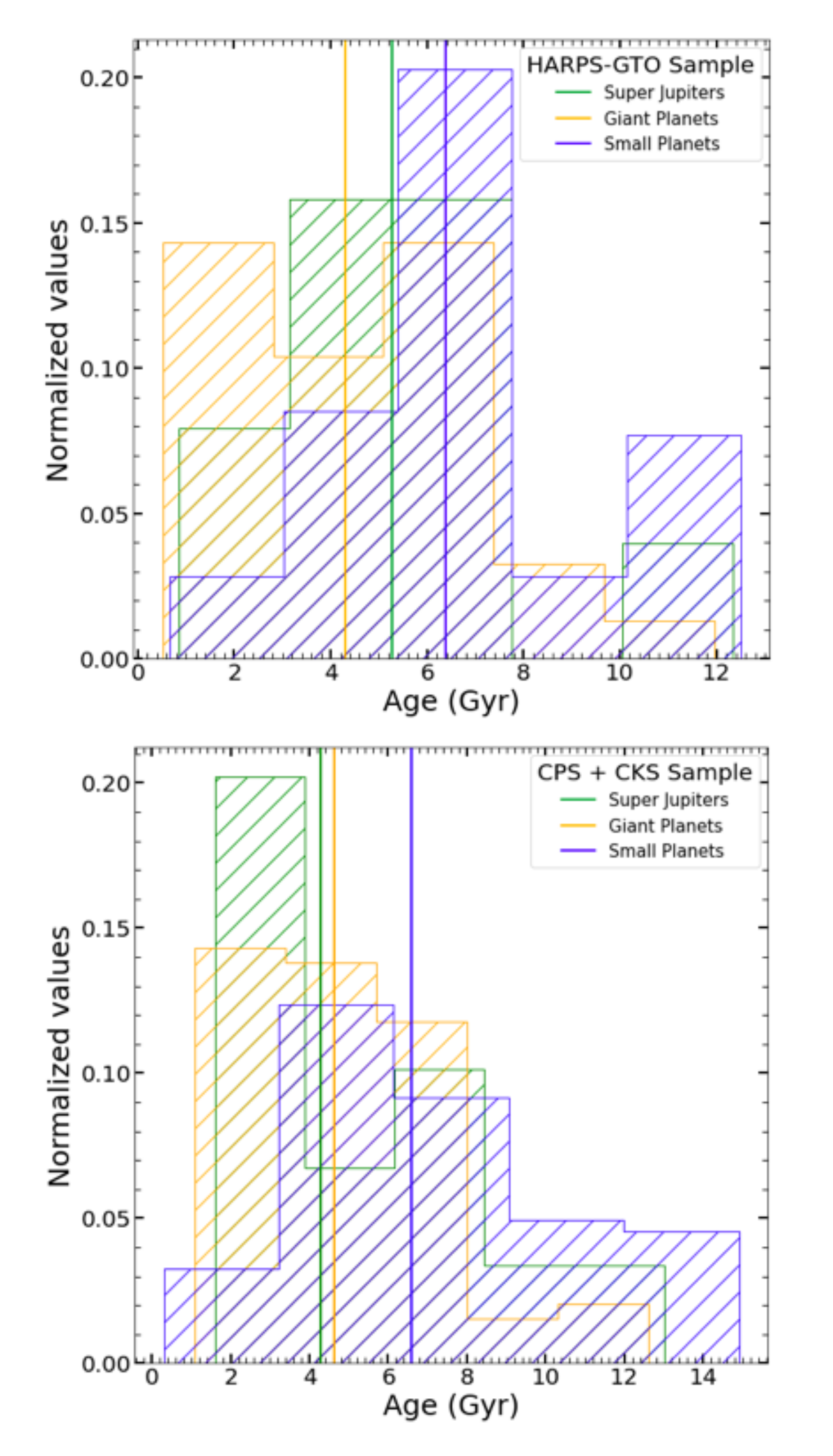}
\caption{Age distribution of planet hosting stars derived using isochrone fitting. The vertical lines represent the median for each of the distribution.}
\label{md}
\end{figure}

For the heavy elements, the trends with planet mass can also be explained by the hypothesis discussed above. In the case of Eu, which is a r-process element, is largely formed through neutron-star mergers \citep{2017Sci...358.1570D,2018ApJ...855...99C}. These merger events  predate the time scales of SNe~Ia \citep{2020A&A...634L...2S,2021AJ....162..229R}. Further, studies have also shown that Eu abundance increases with age \citep{2015A&A...578A..87S,2019A&A...624A..78D} similar to $\alpha$-elements. In our analysis, we find that Eu elemental abundance decreases as a planet mass increases, a behaviour similar to that of $\alpha$-elements. The decrease of [Eu/Fe] with planet mass further strengthens our results and supports our hypothesis that exoplanet host stars with planet mass $>$0.3$M_{J}$ could indeed be younger than SP hosts.

\begin{figure*}[!t]
\includegraphics[width=2\columnwidth]{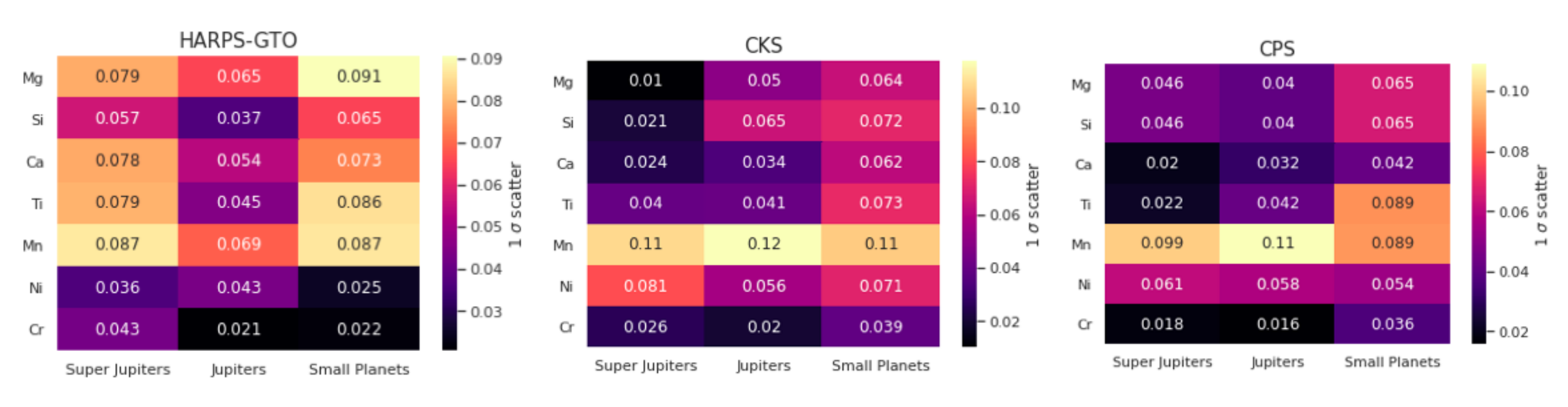}
\caption{Heat-map of 1$\sigma$ scatter from Figure~\ref{al} and Figure~\ref{ir} for the abundance dispersion of the host stars of small planets, jupiters and super-jupiters. A linear de-trend was applied to the original abundance data to compute the 1$\sigma$ scatter for each mass-bin.}
\label{hm}
\end{figure*}

The s-process elements are primarily produced in low-mass AGB stars; thus, their contribution is expected to increase with time. The light s-process (Sr, Y, Zr) elements also show trends similar to iron-peak elements. Their trends for chemical abundances have a negative correlation with age ($\leq 8\: Gyr$) as shown by other studies \citep[e.g.][]{{2016A&A...586A..49B,2019A&A...624A..78D}}, which is expected as their production timeline is similar to the iron-peak elements. For heavy s-process element Ba, we see a positive slope with planet mass, but the trends are opposite for Nd, with Nd showing a strong negative trend with planet mass similar to Eu. This is because even though Nd is considered as a heavy s-process element, only about 56 per cent of it is formed via the s-process \citep{1999ApJ...525..886A, Bisterzo_2016}. The remaining Nd is produced by the r-process, predating the timescales scales of the s-process. Thus, in light of GCE, the $\alpha$-element abundance seems as a good proxy for the planet mass. One possible implication of this finding could be that star hosting small planets have been forming through all epochs while the formation of star hosting giant planets and super-jupiters happened in later epoch when the ISM was sufficiently enriched by the iron and iron-peak elements.

\subsection{Independent age analysis}
\label{AGE}
The negative correlation  between $\alpha$-elements and planet mass presented in Section~\ref{xfe} indicates that stellar hosts of giant planets are probably younger. To corroborate our result, we took the independent age estimates of the HARPS-GTO, CPS and CKS sample from \cite{2019A&A...624A..78D}, \cite{2016ApJS..225...32B} and \cite{2018ApJS..237...38B}, respectively. In these studies, the stellar ages were determined using the isochrone fitting technique requiring effective temperature ($T_{\mathrm{eff}}$) and luminosity ($L$) which were  obtained from photometric and spectroscopic studies. For the HARPS-GTO sample, the ages were estimated using DR2 parallaxes and PARSEC isochrones \citep{2012MNRAS.427..127B}, while for the CPS and CKS sample the ages were determined by Yonsei - Yale isochrones \citep{2004ApJS..155..667D} and Dartmouth Stellar Evolution isochrones \citep{2008ApJS..178...89D}. Although, the individual age estimates will vary depending on the choice of model, it will not  impact the underlying statistical trends, as these models do not have any significant systematics \citep{2019A&A...624A..78D}.

The age distribution of our samples divided into three mass bins is shown in Figure~\ref{md}. We see that the host stars of giant  planets ($M_{P}>0.3M_{J}$) and super-jupiters are younger compared to stars hosting smaller planets. For example, the median age (in Gyr) of SP, GP and SJ hosts  is 6.40, 4.30 and 5.23 for the HARPS-GTO sample and 6.58, 4.62 and 4.3 for the CPS and CKS  sample. For planet hosting stars in HARPS-GTO sample, \cite{2019A&A...624A..78D}  (see their Figure 7) have found  a positive correlation between  [$\alpha$/Fe] abundances and stellar age. This combined with the decrease of negative slope between [$\alpha$/Fe] and planet mass found in this work, further lends independent support to our inference that stellar systems harbouring massive planets could be younger.

Further, within the giant planet population, there is an observed paucity of hot-jupiters around old stars. To explain the dearth of old stars hosting massive and hot planets, \cite{2019AJ....158..190H} have argued that the tidal interaction between the host star and the planet can cause the planet to spiral into the star. Older stars will lose their hot Jupiters if this tidal infall timescale is relatively short. However, the tidal infall timescale (based on Equation 4 in \citealt{2019AJ....158..190H}) for a Jupiter-like planet around a Sun-like star for orbital periods $> \sim$7 days can be as long as the main-sequence lifetime of these stars. The tidal infall timescales are even longer for planets less massive than Jupiter. Only giant planets with $M_P >$ 2 $M_J$ and orbital period $< 5$ days have tidal infall timescale much shorter than 1~Gyr. Hence tidal infall might be playing a key role, but it alone cannot explain the lack of hot giant planets around older stars (also see Narang et al., 2022 under rev).

Abundance scatter or dispersion is another measure of implicit spread in stellar ages. With the chemical evolution of the galaxy, the dispersion in elemental abundance, especially the [$\alpha$/Fe] is expected to increase. We refer to the abundance spread seen in Figure~\ref{al} and Figure~\ref{ir}. This spread is further quantified in Figure~\ref{hm} where we show the heat-maps for the 1$\sigma$ scatter in the abundance distribution of the stellar hosts. From Figure~\ref{hm}, the 1$\sigma$ scatter for $\alpha$-element abundances is much more pronounced for the small planet-hosting stars than the Jupiter and super-jupiter hosts. Also, the scatter in the iron-peak elements, except for Mn, is relatively smaller than $\alpha$-elements in all three samples. The large scatter in the $\alpha$-element abundances implies a large dispersion in ages of stars hosting small planets. Which means, small planets started forming early in our galaxy when [$\alpha$/Fe] was high and continued to form in later generations of stars when [$\alpha$/Fe] has declined. On the contrary, abundance distribution of stars hosting Jupiter-analogs and super-jupiters has small scatter and hence similar age, indicating the massive planets belong to a later generation of stars, represented by overall low [$\alpha$/Fe] and increased iron-peak element abundance. Note that the overall scatter seen in Figure~\ref{al} and Figure~\ref{ir} is much larger than the error in abundance determination of individual stars which is typically, 0.02-0.05 dex.

\begin{figure*}[!t]
\centering
\includegraphics[width=1.8\columnwidth]{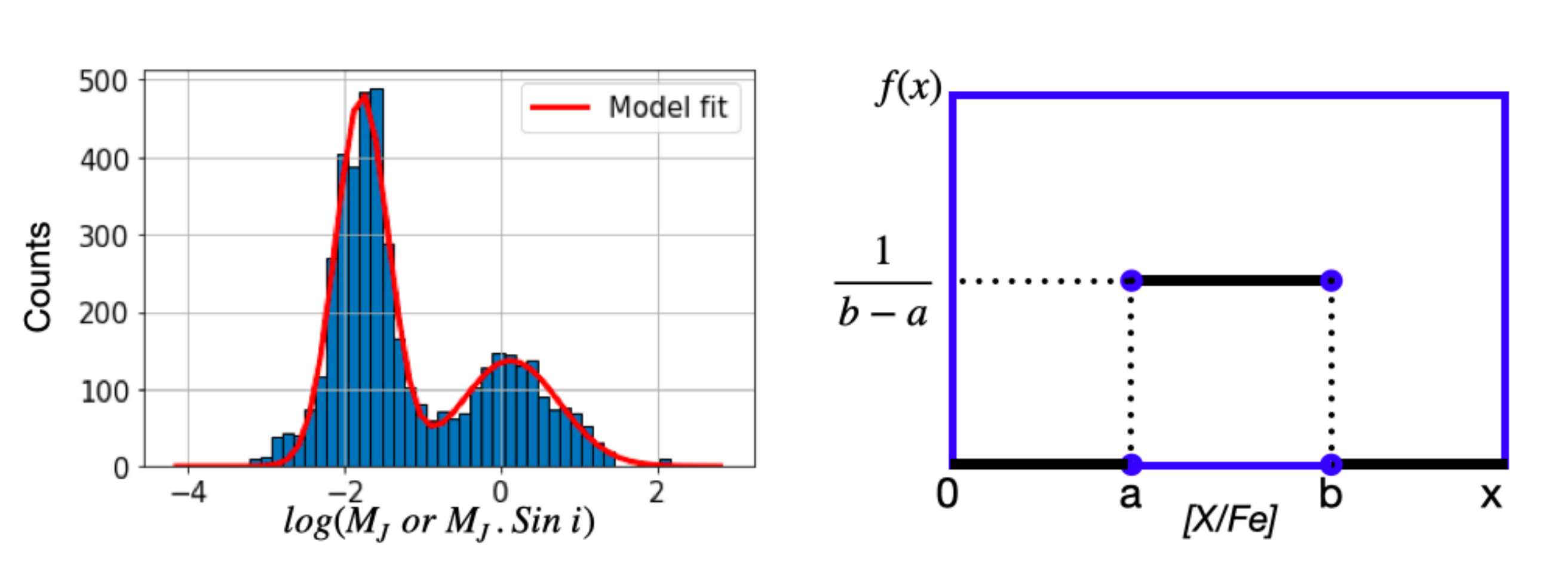}
\caption{\textbf{left}: The observed distribution of exoplanet mass (in log scale) taken from NASA exoplanet archive. \\ \textbf{Right}: A uniform distribution assumed for $\alpha$-element abundances of planet hosting stars.}
\label{dis}
\end{figure*}
\subsection{GCE and formation of giant planets}
\label{CE}
Results from Section~\ref{xfe} clearly establish a link between the abundance of iron and  iron-peak elements and the giant planets. Among existing theories, gravitational instability is proposed as a preferred method for the giant planets formation  beyond the snow line \citep{1997Sci...276.1836B,2002Sci...298.1756M}. Additionally, \cite{kra10} showed  that for a runaway gravitational instability of the disk to happen, the  gas-cooling time has to be shorter than the Keplerian shearing time scale. One could speculate that at large distance ($\sim$10s AU) from the star, the radiative losses from the metals in protoplanetary disk could possibly contribute to the cooling of the gas  during the nascent stage. However, scores of directly imaged planets and brown dwarfs found in wider orbits have not shown any marked dependence on their host star metallicity \citep{2021AJ....161..114S}.

The close by gas giants detected by transit and RV surveys  are expected to form via core accretion process \citep{1996Icar..124...62P, 2007ApJ...662.1282M,2016SSRv..205...41B,2018MNRAS.480.2206O,2022arXiv220309759D}. The galactic chemical evolution trends for $\alpha$ and iron-peak elements in Appendix~\ref{app} (Figure~\ref{alfe}-\ref{lspfe}) shows that in the region [Fe/H] $>$ -0.5, [$\alpha$/Fe] decreases with the enrichment of [Fe/H], but for iron-peak elements, the trends are mostly flat or increasing with [Fe/H]. Therefore, with the enrichment of Fe in ISM, the content of iron-peak elements scales much faster with Fe compared to the $\alpha$-peak abundances. The fact that the gas giants are known to be formed from a metal-rich protoplanetary disk is a natural consequence of the large addition of iron-peak elements.

To form a gas giant via the core accretion mechanism, two significant steps must be followed. First is the formation of a solid mass embryo with a mass of about $\sim$ 10$M_{\oplus}$ in the protoplanetary disk by numerous collisions and coagulation of the planetesimals (for more comprehensive details, please refer \citet{2022arXiv220309759D}). The second is the rapid accretion of gas from the protoplanetary disk before the gas and dust are completely dissipated \citep{2003ApJ...598L..55R,2016SSRv..205...41B,2022arXiv220309759D}. To accrete and form a gaseous envelope around the solid core, the core must grow relatively faster (3-10 Myr) \citep{2007ApJ...662.1282M,2012MNRAS.427.2597A,2021A&A...656A..70E,2022arXiv220309759D}. Whether a proto-planet will end up like a rocky planet or a gas giant will depend on the amount of material present in the proto-planetary disk \citep{2005A&A...434..343A}. The gas-giant planet formation requires the core to build faster to outdo the gas dissipation rate so that the gas is not entirely depleted by the time the massive core ($\sim$ 10$M_{\oplus}$) is formed. Although protoplanetary disks were massive during the early phase of GCE, the refractory $\alpha$-elements alone would have contributed to the formation of the core. By the time the core formed, most of the gas in the disk would dissipate, leading to a preferential formation of rocky planets. However, as the galaxy chemically evolved, the ISM was enriched in both $\alpha$ and iron-peak elements coming from the SNe Ia. This additional enrichment would propel the growth of grains, pebbles, planetesimals and finally, the core \citep{2021A&A...656A..70E,2022arXiv220309759D}. Since the chemically enriched material fuels the growth of the core quickly reaching the critical mass, the disk will have sufficient gas left to accrete onto the surface of the core to form the gas giants \citep{2007ApJ...662.1282M,2012MNRAS.427.2597A,2021A&A...656A..70E,2022arXiv220309759D}. Thus, the delayed enrichment of ISM by SNe Ia created pathways for the formation of gas giants, which seems consistent with the core accretion process.

\subsection{Biases and statistical validity of our results}
\label{MM} 
Exoplanet host stars selected from various transits and RV surveys could suffer from different detection and selection biases. The transit method, for example, is primarily sensitive to planets orbiting close to the star with near edge on configuration. The radial velocity method, on the other hand, is suitable to detect giant planets at large orbital distances (see e.g., Figure~\ref{rvtrans}). Since the RV precision is adversely effected by stellar activity and line broadening mechanisms,  the highly active and fast rotating stars are usually excluded from the RV surveys, keeping the focus largely on the main sequence dwarfs. However, to our knowledge no study exists which links the chemical composition of stars to stellar activity and/or rotation. Therefore, it is very unlikely that possible biases in RV/transit search would impact the chemical analysis of the exoplanet hosting stars. As reported in section~\ref{xfe}, the stellar abundance trends  with planet mass are similar and consistent with GCE for all three samples, regardless of the search method.

Further to ensure our results are not biased due to low-number statistics or random correlations in the abundances, a Monte Carlo test was carried out . We attempted to reproduce the correlations obtained in Section~\ref{xfe} by using simulated planet mass and abundances. Since the trends for $\alpha$-elements were most robust, we used them as a case study for this analysis. For our simulations, we constructed a bimodal function describing the observed planet mass distribution as shown in the left-panel of Figure~\ref{dis}. The apparent mass distribution of confirmed exoplanets has two peaks, - one near 0.01~$M_{J}$ and another at 1~$M_{J}$. The intrinsic mass distribution of exoplanets could be different from the apparent mass distribution, but we discount any selection effects since they are hard to quantify \citep{Malhotra_2015}.  

\begin{figure*}[!t]
\centering
\includegraphics[width=2\columnwidth]{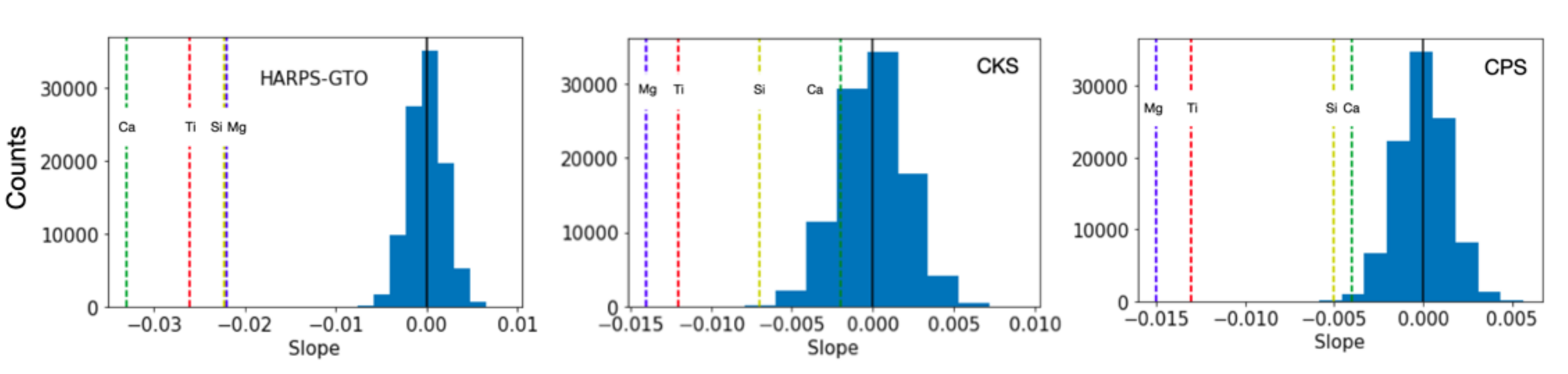}
\caption{The distribution of numerical slope (between $\alpha$-element  abundances of host star versus planet mass) generated from Monte-Carlo simulations for the HARPS-GTO, CKS and CPS samples. The solid black line represents the mean of the distribution and color dotted lines represent the measured slope from Figure~\ref{al}.}
\label{ap2}
\end{figure*}

For abundances, we assumed a uniform distribution, bound by a rectangular window function shown in the right-panel of Figure~\ref{dis}. The lower and upper bounds for the abundance distributions in each sample are taken at 3$\sigma$ cutoff on either side of the observed distribution mean of the averaged $\alpha$-abundances plotted in the last row of Figure~\ref{al}. 

For each simulation run, we randomly draw 500 samples from the assumed distributions of planet mass and abundances. We then calculate the slope of the best fit line between the planet mass and abundances. The goal is to check how often the observed slope is reproduced in a reasonably large numerical experiment, with underlying abundance distribution assumed to be uniform. The simulation was repeated 100,000 times, and the final histograms of slopes are shown in Figure~\ref{ap2}. For the HARPS-GTO sample, the observed slopes for all the elements are significantly far from the mean of the numerical slopes, suggesting that observed trends are highly improbable due to chance outcome. In fact, none of the trials produced result that matched the observed slopes,  thus rendering the probability of observed trends arising from random occurrence extremely low ($\leqslant 10^{-5}$).  Similarly, for the CKS and CPS samples, the observed slopes for Mg, Si and Ti are also significantly away  from the mean of the numerical distribution, with Ca as exception which is only about  1~$\sigma$ (CKS) and 3~$\sigma$ (CPS) away. Generally, this analysis suggests that the observed  $\alpha$-element trends obtained in Section \ref{xfe} cannot be generated by simple randomness, and these trends are also not due to low-number statistics. Therefore, \textit{planet-abundance} pattern observed in a finite sample of exoplanet host stars must be a correct manifestation of the underlying population.  

\section{Summary and conclusions} 
\label{s5}
Exoplanet properties are intimately connected to the properties of their stellar hosts. In this work we studied the chemical abundances of planets hosting stars for planets in different mass bins. We analyzed the abundances of 17 elements belonging to different classes based on their formation mechanism and evolution of chemical history of the galaxy. We used data from well known exoplanet search programs, namely, HARPS-GTO, CKS and CPS and planetary mass from NASA exoplanet catalog. Our analysis includes 968 planet-hosting stars, which are discovered by both transit and radial velocity method. Here, we present a summary of our results :

\begin{enumerate}

\item We find that for all the $\alpha$-elements, which are mainly produced in SNe II, there is an unambiguous negative slope with planet mass for all the three samples used in this study, showing that stars hosting small planets are  clearly $\alpha$-rich compared to stars harboring giant planets and super-jupiters.

\item We find a positive correlation for Mn and near-zero correlation for the iron-peak elements for almost all the cases with planet mass. Since iron-peak elements are primarily formed during SNe Ia, and followed the same scaling as iron, their surface composition [X/Fe] in stellar hosts remains mostly the same, regardless of the planet mass.  

\item For the r-process element, Eu, which is mainly produced by neutron star mergers, happened at much earlier stages of GCE (earlier than SNe Ia), and thus, [Eu/Fe] versus planet mass trend is similar to $\alpha$-elements.

\item The s-process elements are primarily produced in AGB stars and are formed at much later stages of the GCE (after SNe Ia enrichment). We expected their trends to follow iron-peak elements. However, we find that Nd shows a significant negative trends with planet mass. This could be because a significant amount of Nd is produced by r-process.

\item Our abundance analysis of exoplanet host stars shows two specific trends with planet mass; --a distinct negative slope for alpha-elements including Eu and a near-zero slope for most iron-peak elements. Seen in the context of GCE, these results imply that stellar systems with small planets may have started forming early in the evolutionary history of our galaxy, whereas, the emergence of high-mass planetary systems had to wait until the ISM was sufficiently enriched.

\item To validate our findings, we compare the stellar ages estimated from the isochrone fitting. Our independent age analysis also shows that host stars of massive gas giant planets are indeed statistically younger than the stars hosting low-mass planets. 

\item  Compared to their low-mass counterparts, we also find a relatively small scatter in the abundance distribution of stellar hosts of high-mass planets. This is compatible with the younger age and temporal offset in the formation scenario of jupiters and super-jupiters. 

\item Our sample of multiplanetary systems hosting at least one low-mass and one high-mass planetary companion, do not show any correlation between [$\alpha$/Fe] and planet mass. In addition, their overall [$\alpha$/Fe] abundance  across all three mass-bins is also lower. This too, suggests the possibility that such multiplanetary systems are younger. 

\end{enumerate}

In conclusion, we have analysed the elemental abundances of a large sample (968) of planet-hosting stars, connecting the planet formation process to the evolution of the chemical enrichment of the ISM. The detailed abundances of exoplanet host stars are largely consistent with the galactic chemical evolution. From the observed trends between stellar abundances and planet mass, we conclude that the low-mass planets may have been formed during all epoch of star formation, while the giant planets are formed around chemically enriched stars that are relatively young. 

To strengthen these findings, the future high-resolution spectroscopic surveys should target a bigger sample of exoplanet hosting stars, determining their chemical abundances uniformly and homogeneously. More theoretical and experimental work is required to further understand the importance of chemical abundance, specially, the role that iron-peak elements play in the formation and growth of dust grains, pebbles and planetesimals in the astrophysical environment.

\section*{ACKNOWLEDGMENTS}
We would like to thank the anonymous referee for the useful comments and suggestion which helped us in improving  the manuscript. This work has made use of (a) the NASA Exoplanet Database, which is run by the California Institute of Technology under an Exoplanet Exploration Program contract with the National Aeronautics and Space Administration and (b) the European Space Agency (ESA) space mission Gaia, the data from which were processed by the Gaia Data Processing and Analysis Consortium (DPAC) (c) the exoplanet.eu database maintained developed and maintained by the exoplanet TEAM. C. Swastik would like to thank E. Delgado Mena for providing us with the list of exoplanet hosting stars for the HARPS-GTO sample. C. Swastik would also like to thank P.P Goswami and Pallavi Saraf for the insightful discussion on formation of heavy elements.

\software{Astropy \citep{astropy:2013}, Scikit-learn \citep{scikit-learn}, Topcat \citep{2005ASPC..347...29T}, Scipy \citep{2020NatMe..17..261V}, Matplotlib \citep{4160265}, numpy \citep{2020Natur.585..357H}}

\bibliography{biblio}{}
\bibliographystyle{aasjournal}

\newpage
\appendix
\section{Elemental abundances [X/Fe] as a function of [Fe/H]}
The [X/Fe] versus [Fe/H] trends for the sample of exoplanet host stars studied in this paper are shown in Figures 12-14. In the case of $\alpha$-elements we see the [X/Fe] trend to be decreasing with [Fe/H], while for iron-peak elements, we see a variety of trends with [Fe/H]. The overall trends seen in these figures are consistent with the standard GCE model. The planet-hosting stars primarily lie in the metal-rich regime, as seen below.
\label{app}

\begin{figure*}[h]
\centering
\includegraphics[width=0.82\columnwidth]{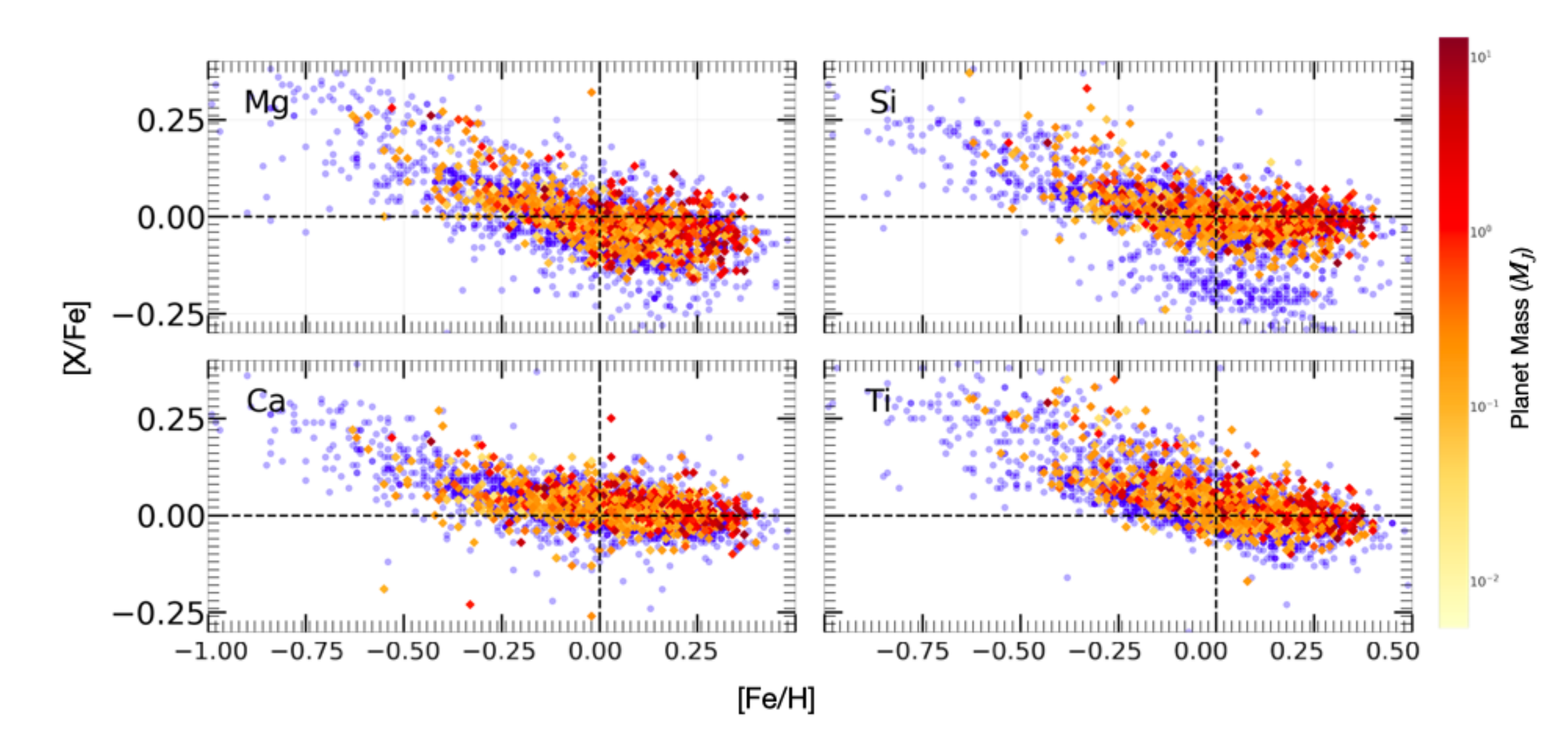}
\caption{Abundance ratios [X/Fe] vs [Fe/H] for $\alpha$-elements for stars belonging to all the three samples : HARPS-GTO, CKS and CPS. The blue dots represent stars without planets while the colorbar represents stars hosting planets of different mass.}\label{comp}
\label{alfe}
\end{figure*}

\begin{figure*}[h]
\centering
\includegraphics[width=0.8\columnwidth]{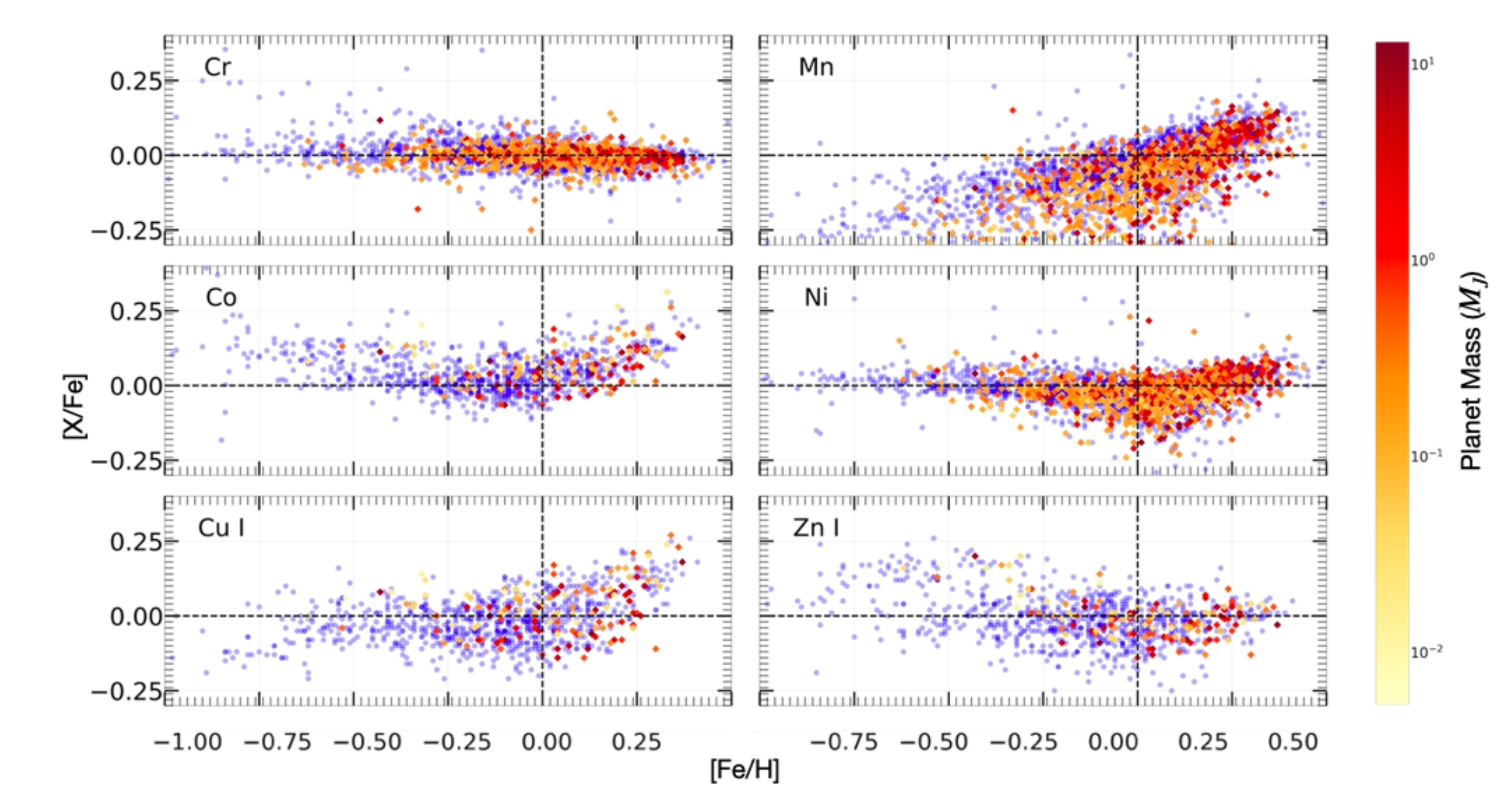}
\caption{Abundance ratios [X/Fe] vs [Fe/H] for iron-peak elements for stars belonging to all the three samples : HARPS-GTO, CKS and CPS. The blue dots represent stars without planets while the colorbar represents stars hosting planets of different mass.}
\label{irfe}
\end{figure*}
\begin{figure*}[h]
\centering
\includegraphics[width=0.63\columnwidth]{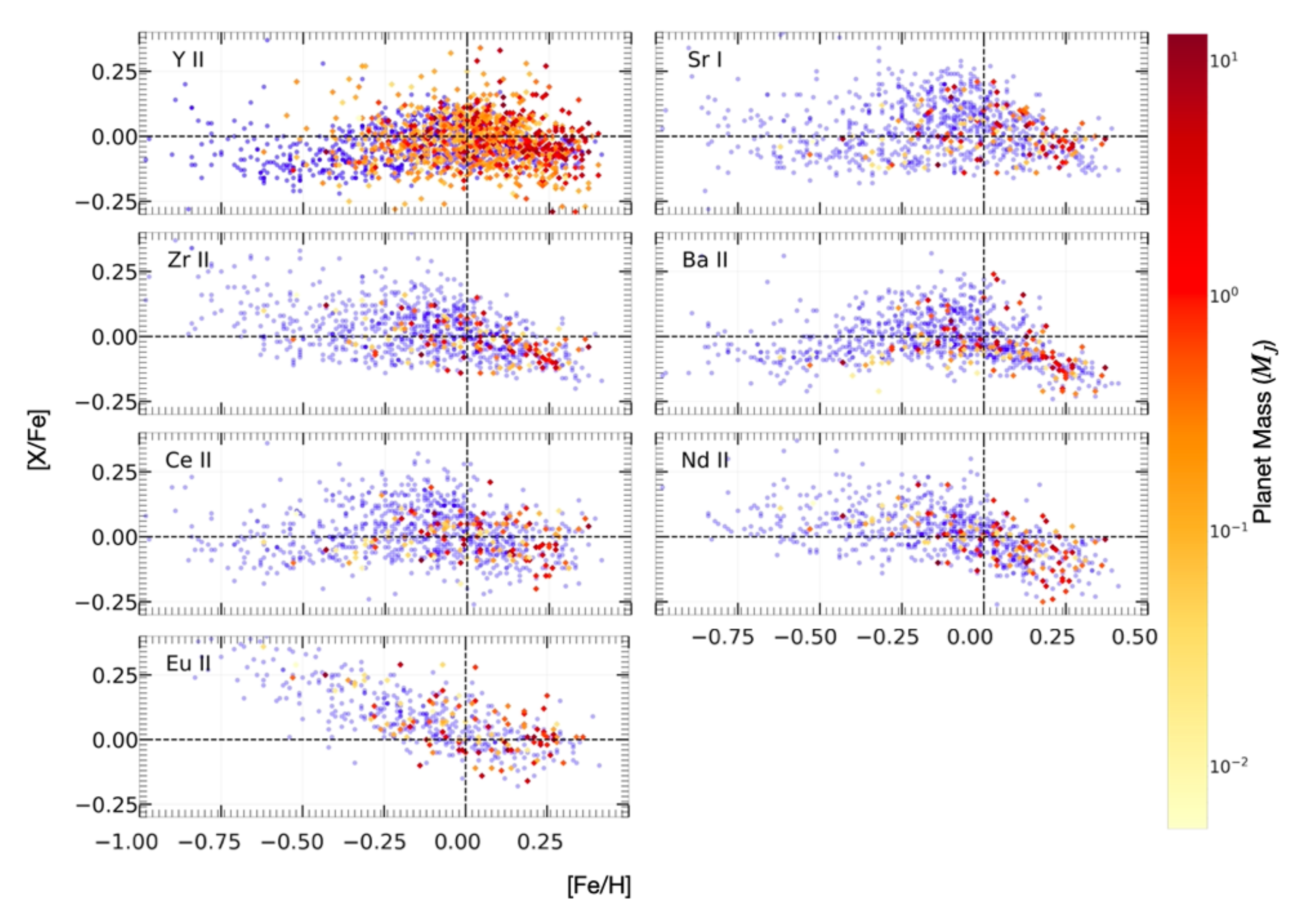}
\caption{Abundance ratios [X/Fe] vs [Fe/H] for heavy elements (A$>$30) for stars belonging to all the three samples : HARPS-GTO, CKS and CPS. The blue dots represent stars without planets while the colorbar represents stars hosting planets of different mass.}
\label{lspfe}
\end{figure*}
\section{$\alpha$-element abundance for the multi-planetary systems.}
\label{apb}
\begin{figure*}[h]
\centering
\includegraphics[width=0.63\columnwidth]{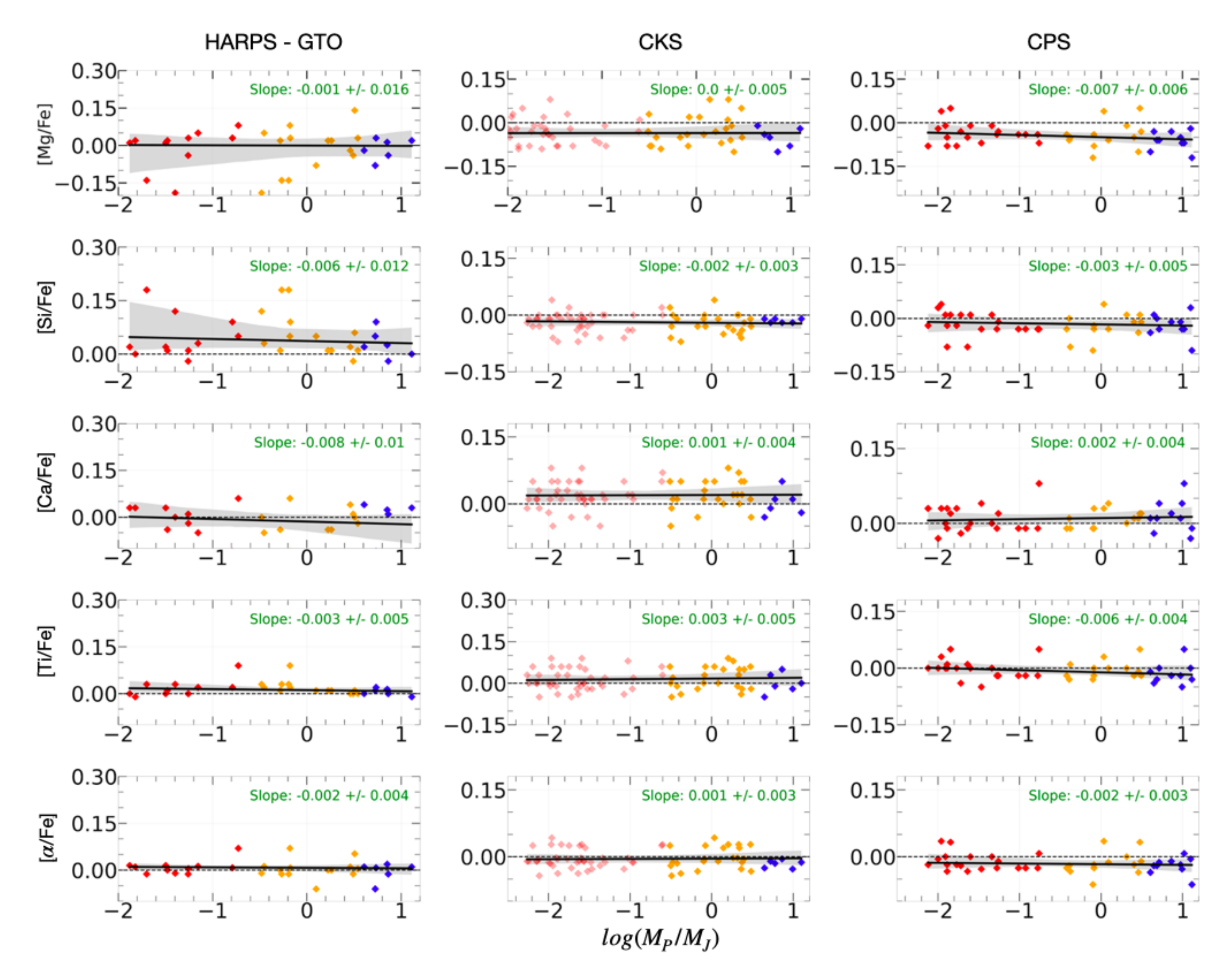}
\caption{$\alpha$-element abundances as a function of planet mass for the multiplanetary systems which host at least one planet each in low-mass and Jupiter and/or super-jupiter mass regimes. The last row is the arithmetic mean of the $\alpha$-element abundance from the  above four rows.}
\label{mulp}
\end{figure*}

For the subsample of multi-planetary systems excluded from the analysis in Section~\ref{xfe}, we do not find noticeable correlation between the $\alpha$-element abundance and the planet mass (see Figure~\ref{mulp}). This indicates the multiplanetary systems that hosts at least one giant or a super-giant, are also recently formed.  
\end{document}